\documentclass[a4paper, 11pt, margin=1in]{article}
\usepackage{amsmath,amsthm,amssymb,mathrsfs,graphicx,url}
\usepackage{fullpage}
\usepackage{multirow}
\usepackage[usenames]{color}
\usepackage{ifthen}
\usepackage{setspace}
\usepackage{caption}
\usepackage{subfigure}
\usepackage{float}
\usepackage{graphicx}
\usepackage{algorithm}
\usepackage{bm}
\usepackage[noend]{algpseudocode}
\usepackage{enumitem}
\usepackage{amsfonts}

\usepackage{natbib}
\usepackage{authblk}

\makeatletter
\def\BState{\State\hskip-\ALG@thistlm}
\makeatother

\def\keywords{\vspace{.5em}
{\textbf{Keywords}:\,\relax%
}}

\theoremstyle{plain}

\theoremstyle{remark}
\newtheorem{remark}{Remark}

\theoremstyle{definition}
\newtheorem{defn}{Definition}

\theoremstyle{remark}

\theoremstyle{definition}

\newcommand{\prob}{\mathsf{P}}

\renewcommand{\phi}{\varphi}

\title{Modeling Binary Time Series Using Gaussian Processes With Application to Predicting Sleep States}
\author{Xu Gao$^{1}$, 
	Babak Shahbaba$^{1}$\, 
	Hernando Ombao$^{1,2,3}$ \\
	$^{1}$Department of Statistics, University of California, Irvine, California, U.S.A. \\
	$^{2}$Department of Cognitive Sciences, University of California, Irvine, California, U.S.A.\\
	$^{3}$Program on Applied Mathematics \& Computational Science, \\
	King Abdullah University of Science and Technology, Saudi Arabia\\
\tt{xgao2@uci.edu \ \ babaks@uci.edu \ \ hernando.ombao@kaust.edu.sa}
}
\date{}

\begin{document}

\maketitle
\begin{abstract}
Motivated by the problem of predicting sleep states, we develop a mixed effects model for binary time series with a stochastic component represented by a Gaussian process. The fixed component captures the effects of covariates on the binary-valued response. The Gaussian process captures the residual variations in the binary response that are not explained by covariates and past realizations. We develop a frequentist modeling framework that provides efficient inference and more accurate predictions. Results demonstrate the advantages of improved prediction rates over existing approaches such as logistic regression, generalized additive mixed model, models for ordinal data, gradient boosting, 
decision tree and random forest. 
%Furthermore, the procedure yields efficient inference on the coefficients of covariates. 
Using our proposed model, we show that previous sleep state and heart rates are significant 
predictors for future sleep states. Simulation studies also show that our proposed method is promising and robust. To handle computational complexity, we utilize Laplace approximation, golden section search and successive parabolic interpolation. With this paper, we also submit an R-package ({\tt HIBITS}) that implements the 
proposed procedure. 
\end{abstract}


\section{Introduction}

\begin{figure}[th!] \centering
	\begin{tabular}{cc}
		\includegraphics[width=.5\textwidth]{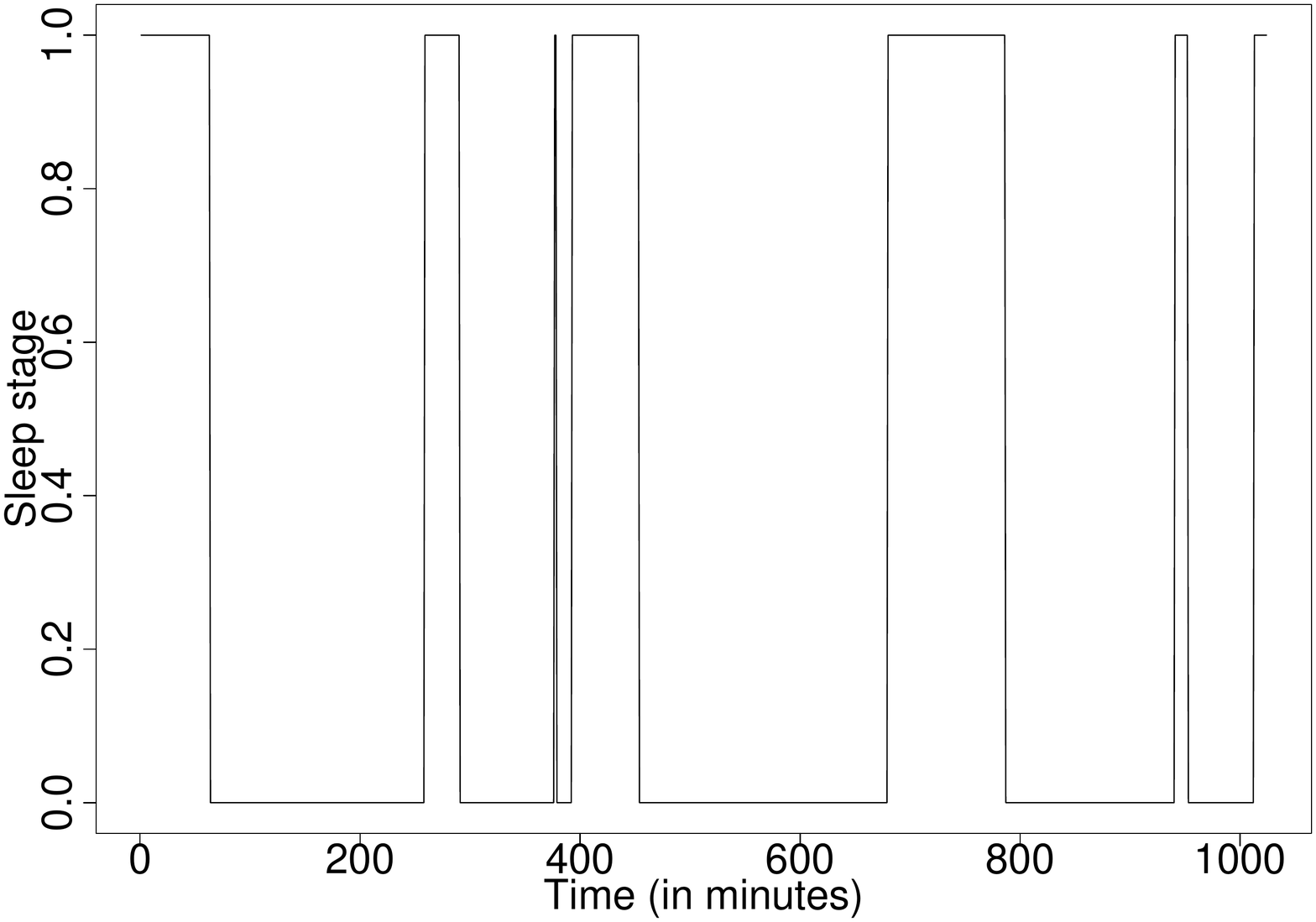} &
		\includegraphics[width=.5\textwidth]{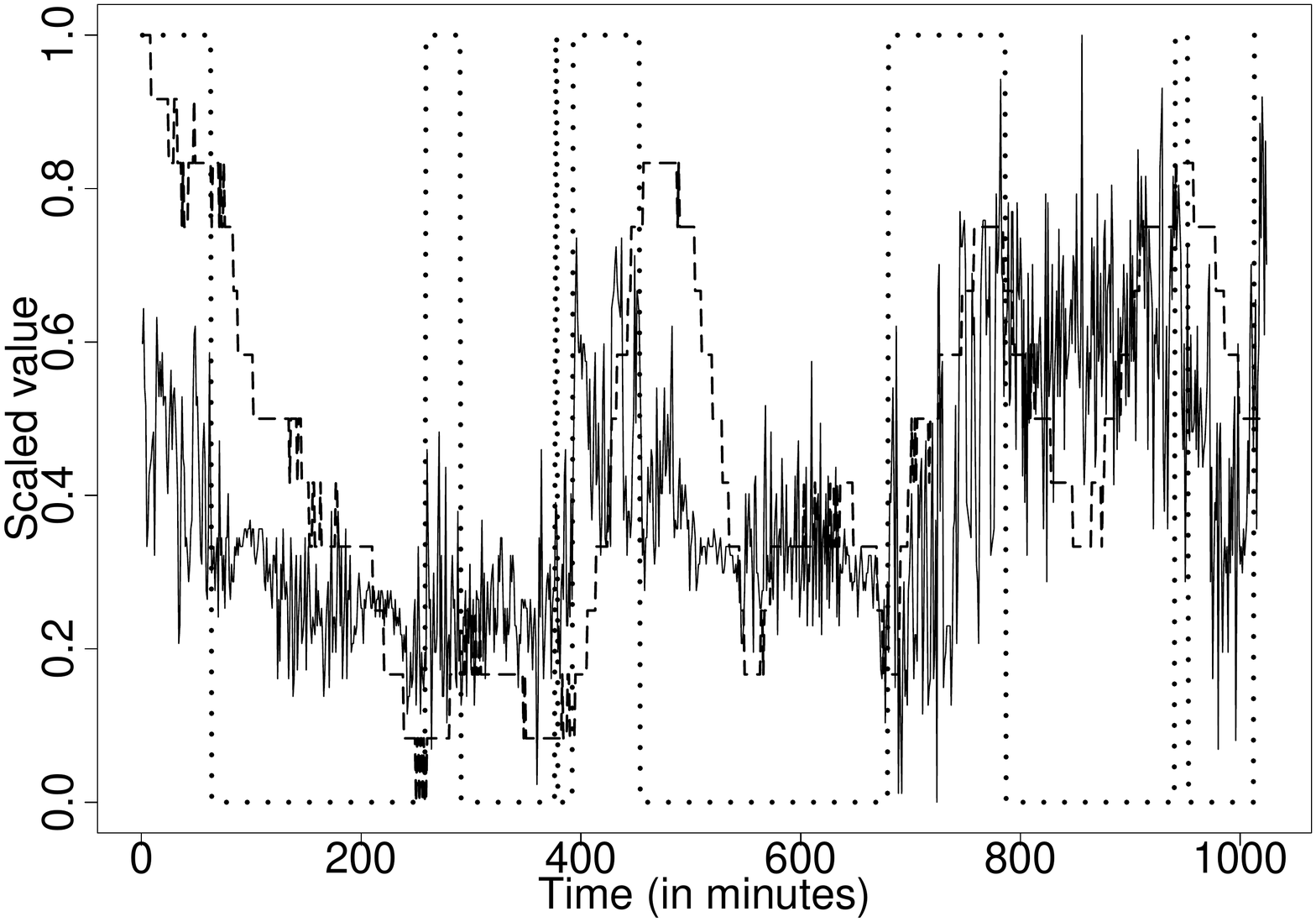} \\                
%		Sleep state & Overlaid plot (scaled heart rate in black, \\
%		&temperature in red and sleep state in green) 
	\end{tabular}
	\caption{Left: sleep state. Right: sleep state plot (dotted line) overlaid by scaled 
	heart rate (solid line)	and body temperature (dashed line) time plots.}
	\label{expl2}
\end{figure}

The American Academy of Sleep Medicine indicates that humans go through several cycles during sleep with each cycle comprised of different stages. It is important to study sleep in humans because the lack of sleep is associated with 
psychiatric diseases (e.g. depression and ADHD) and  chronic diseases (e.g. diabetes, heart disease and hypertension). In particular, understanding sleep state (asleep versus awake) and uncovering its latent pattern play a critical role in people's daily routine. For example, many young mothers wonder if their infant's sleep state can be predicted in advance; physicians are interested in forecasting their patient's anesthesia level/sleep state for surgery. The goal of this paper is to develop statistical 
inference for studying changes in the sleep state (in particular, asleep versus awake) and the potential roles of covariates such as heart rate, respiration rate and body temperature on sleep states. A plot of the sleep states and the exogenous time series of heart rate and temperature, given in Figure~(\ref{expl2}), suggest a lead-lag depenence between sleep states and the exogenous time series. In this paper, we develop a model that formally tests for these lead-lag dependence and predict future sleep states.  

Various approaches have been proposed to model and predict sleep states. Since sleep states can be measured sequentially in time during an experiments or in observational studies, typical strategies for modeling categorical time series have been implemented. \cite{caiado2006periodogram} introduced new measurements in classifying time series based on periodograms. \cite{maharaj2002comparison} put forward a framework of comparing time series in frequency domain. Wavelet based clustering method was also introduced by \cite{maharaj2010wavelet}. \citet{jacobs1978discrete} proposed a discrete autoregressive-moving average (DARMA) model by utilizing probabilistic mixtures. Recently, \cite{gao2017fisher} proposed efficient statistical inference on logistic autoregressive model, which is a widely used example of binary time series. A comprehensive modeling framework based on generalized linear models and partial likelihood inference have been developed in \citet{fokianos2002regression} and \citet{fokianos2003regression}. \citet{fokianos1998prediction} extended the partial likelihood analysis to non-stationary categorical time series including stochastic time dependent covariates. With the Markovian structure, \citet{meyn2012markov}, \citet{bonney1987logistic} and \citet{keenan1982time} developed inferential procedures based on the conditional likelihood. These previous studies provide inference on binary time series. Their main drawback is that they involve massive computation for high dimensional integrals, which results in poor prediction accuracy. \citet{lindquistJASA09} introduced a logistic regression model with functional predictors and extended it to generalized linear model. Their substantial work was superior in detecting sensitive and interpretable time points that were most predictive to the response. However, when it is applied to this study, the drawbacks are: (1) the Brownian motion assumption is unlikely to be satisfied in practice because the covariates in this study hardly have the property of increment independence; (2) the influence of covariates on responses  is assumed to spread across the entire trajectory and hence implies the non-existence of ``sensitive time points"; (3) prediction of the time series is not developed, which could be a serious limitation for this project since we are also interested in such predictions. 

From the view of the machine learning community, typical classification methodologies such as decision tree, random forest and strategies such as boosting can also be used for predicting sleep states. Although such approaches are able to achieve predictions with high accuracy, the major drawback is that they give very little guidance to sleep researchers who are measuring the impact of previous heart rate, temperature and respiratory rate on future 
sleep state. In this paper, we develop a statistical model that can provide us simultaneously 
with convincing inference and interpretation at the same time produce prediction accuracy that is higher than that achieved by typical machine learning classification approaches. 
%It is quite common that observations are measured sequentially in time during the procedure of experiments or observational studies. Categorical time series such as binary data arise in a great number of applications \cite{tsa} \cite{bts} \cite{eps} . A variety of different strategies have been proposed for modeling categorical time series. To name a few, Jacobs and Lewis proposed discrete autoregressive-moving average (DARMA) model by utilizing probabilistic mixtures \cite{dts}. Kedem considered means of regression based on generalized linear models and partial likelihood inference \cite{kbf} \cite{fk} . Different link functions have also been discussed in \cite{fk}. Fokianos and Kedem's work in \cite{pcn} extended the partial likelihood analysis to non-stationary categorical time series including stochastic time dependent covariates. Keenan modeled binary data by a underlying Gaussian first-order autoregressive process and different estimation procedure were discussed in \cite{ats}. (Hidden) Markov Models have been implemented when the lagged values are instrumental in determining the future values \cite{ms}. The cumulative odds models were widely used in the application for the analysis of ordinal time series \cite{mp}. The mixture transition distribution (MTD) model was first introduced in  \cite{ra} to tackle the problem of exponentially increasing number of free parameters for a Markov Chain. 

This work is inspired by \citet{keenan1982time} which developed a binary time series using a latent strictly stationary process. The focus here is to provide an accurate, interpretable, efficient yet computationally less demanding approach for estimation and prediction. When prior information indicates that a binary time series is determined by a process comprised with fixed and random components, we decompose the unobserved latent process into linear and stochastic effects with different covariates. On stage one, inference on the fixed effects is conducted using maximum likelihood estimation. On stage two, conditioned on the estimated fixed effect, a Gaussian process will be used to represent the random components. Predictions are obtained by combining inference on these two components. In addition, based on the results from these two stages, we use parametric bootstrap samplers from the estimated Gaussian process to obtain the final point and interval estimates of parameters. 

Using the proposed procedure, we can identify the dependence of the endogenous time series (sleep state) on potential covariates (e.g., heart rate and body temperature) by providing the point and interval estimates of the coefficients from linear effects based on the results from the two-stage algorithm. Inference can be easily and directly performed by maximum likelihood using existing software. Moreover, results are  easily interpretable under the framework of generalized linear model. On stage two, 
which is derived from Gaussian process classification strategy, we can predict the sleep state with high accuracy. Laplace approximation was implemented to reduce the computation cost. This work is also inspired by \cite{brillinger1983generalized} which, to the best of our knowledge, is the first to introduce this notion of a Gaussian random effect as random intercept in a logit model. Here, we generalize this by representing 
the random component as a stochastic process rather than just a scalar random variable. 

The main advantages of our proposed approach, which we call the hybrid inference method for binary time series (HIBITS),  are the following: (1) it accounts for the linear and non-linear stochastic effects 
of covariates and endogenous variables on sleep states; (2) it provides efficient 
point and interval estimates of the coefficients from the linear effects while 
maintaining type I error rates; (3) it produces more accurate predictions compared 
to other existing approaches; (4) it is easily implemented with low computational 
cost; and (5) unlike other classification approaches, it gives more 
straightforward interpretation of the results. 

The remainder of this paper is organized as follows. Section~\ref{gaussian} is devoted to brief introduction of Gaussian process and its existing applications in regression and classification. In Section~\ref{twostage}, we develop our proposed methodology and discuss the motivation and the technical derivation of the proposed HIBITS method. A complete algorithm that yields prediction and inference on the coefficients of covariates and endogenous variables is also provided. Model selection strategy is also developed to address application problems. Section~\ref{simulations} presents the simulation results that show the benefits of the proposed method over the existing methods in terms of the significant higher prediction accuracy and narrower confidence intervals. In Section~\ref{realdata}, we apply our proposed model and inference procedure to identify predictors of sleep states and to predict future sleep states.  The results are promising in terms of prediction accuracy at low computational cost and interpretability. Moreover, the proposed method can also be modified when there are missing values. 

\section{Background on Gaussian Processes in Binary Time Series}
\label{gaussian}

\subsection{Gaussian process and regression models}
Gaussian process have been widely developed in spatial-temporal modeling \citep{williams2006gaussian,banerjee2008gaussian,banerjee2014hierarchical,gelfand2005bayesian,quick2013modeling,stein2012interpolation,zhou2015dynamic,vandenberg2015dependent,wang2014modeling}. % In recent years, Gaussian processes were introduced in the context of machine learning when there was an explosion of work regarding ``kernel machines" \citep{williams2006gaussian}.
It provides a framework that can capture the non-linear and stochastic components of exogenous and endogenous variables based on generalized linear models, which makes it useful for modeling binary time series and classification.

The definition of a Gaussian process is as follows. 
\begin{defn}
A stochastic process is a Gaussian process if and only if for every finite set of indices 
$t_1, \cdots, t_k$ in the index set $T$, $\mathbf{x}=({x}_{t_1},\cdots, 
{x}_{t_k})^T$ is a multivariate Gaussian random variable. 
\end{defn}

We will write the Gaussian process $f(\mathbf{x})$ as $f(\mathbf{x})\sim \mathcal{GP}(m(\mathbf{x}), K(\mathbf{x,x'})),$ where $m(\mathbf{x})=\mathbb{E}[f(\mathbf{x})] \quad \text{and}\quad K(\mathbf{x,x'})=\mathbb{E}[(f(\mathbf{x})-m(\mathbf{x}))(f(\mathbf{x'})-m(\mathbf{x'}))].$
Let us now denote the observed data to be  $\{(\mathbf{x_i},y_i), i=1,\cdots, n+n_*\}$, where $\mathbf{x_i}\in \mathbf{R^p}$ and $y_i$ is the response data. We split the dataset into $n$ training points and $n_*$ testing points. Let ($\mathbf{X_{*}, y_{*}}$) represent the testing datasets and ($\mathbf{X, y}$) represent the 
training datasets respectively. 
Define $
\mu=K(\mathbf{X_*},\mathbf{X})K(\mathbf{X},\mathbf{X})^{-1}\mathbf{y}, \Sigma=K(\mathbf{X_*},\mathbf{X_*})- K(\mathbf{X_*},\mathbf{X}) K(\mathbf{X},\mathbf{X})^{-1} K(\mathbf{X},\mathbf{X_*}).
$
It follows that
\begin{equation}\mathbf{y_*}|\mathbf{X},\mathbf{X_*}, \mathbf{y} \sim N(\mu, \Sigma).
\label{1}
\end{equation}
%where 
%$
%\mu=K(\mathbf{X_*},\mathbf{X})K(\mathbf{X},\mathbf{X})^{-1}\mathbf{y}, \Sigma=K(\mathbf{X_*},\mathbf{X_*})- K(\mathbf{X_*},\mathbf{X}) K(\mathbf{X},\mathbf{X})^{-1} K(\mathbf{X},\mathbf{X_*}).
%$
The distribution of the response $\mathbf{y_*}$ can be determined by Equation (\ref{1}). Point estimates, interval estimates and sampling distribution of $\mathbf{y_*}$ can be derived accordingly. %Alternatively, let us consider introducing noise into the model. Without loss of generality, we assume $y=f(\mathbf{x})+\epsilon$, where $f(\mathbf{x})$ follows a Gaussian process as defined above. Following the similar notation, we denote $f_* = f(\mathbf{X_*})$ to account for the testing data. Then, we have $Cov (\mathbf{y})=K(\mathbf{X},\mathbf{X})+\sigma^2I,$ where $\epsilon\sim N(\mathbf{0}, \sigma^2I).$ The joint distribution of $\mathbf{y}$ and ${f_*}$ yields  $$\left(\begin{array}{c}
%\mathbf{y}\\
%{f_*}
%\end{array}\right) \sim N\left(\mathbf{0}, \left(\begin{array}{cc}
%K(\mathbf{X},\mathbf{X} )+\sigma^2I&  K(\mathbf{X},\mathbf{X_*} )\\
%K(\mathbf{X_*},\mathbf{X} )&K(\mathbf{X_*},\mathbf{X_*} )
%\end{array}\right) \right) .$$ 
%we then conclude that 
%\begin{equation}{f_*}|\mathbf{X},\mathbf{X_*}, \mathbf{y} \sim N(\mu_+, \Sigma_+),
%\label{2}
%\end{equation}
%where
%$
%\mu_+=K(\mathbf{X_*},\mathbf{X})[K(\mathbf{X},\mathbf{X})+\sigma^2I]^{-1}\mathbf{y}, \Sigma_+=K(\mathbf{X_*},\mathbf{X_*})- K(\mathbf{X_*},\mathbf{X}) [K(\mathbf{X},\mathbf{X})+\sigma^2I]^{-1} K(\mathbf{X},\mathbf{X_*}).
%$

%Similarly, inference regarding the response $\mathbf{y_*}$ can be made through the distribution in Equation (\ref{2}).
\begin{remark}
On stage two of the proposed method (discussed in Section~\ref{algo}), results in Equation (\ref{1}) will be utilized to achieve the distribution of the stochastic component which captures the variation in the binary time series beyond which are explained by the covariates. 
\end{remark}
\subsection{Gaussian process in modeling binary time series}
\label{gpic}
\subsubsection{Model formulation}
Denote the observed training data as  $\{(\mathbf{x_i},y_i), i=1,\cdots, n\}$, where $y_i \in \{1,0\}$ and $\mathbf{x_i}\in \mathbf{R^p}$. For our data in this paper, $y_i$ denotes the sleep state at time point $i$ and $\mathbf{x_i}$ can be heart rate or body temperature at time point $i$. We define a latent Gaussian process indexed by $\mathbf{x}$ as $f(\mathbf{x)}$. The relationship between $\mathbf{x_i}$ and $y_i$ is characterized by $\prob(y_i=1|\mathbf{x_i})=t(f(\mathbf{x_i})),$ where $t$ is a link function that determines the relation between $\mathbf{x}$ and the probability of the sleep state. To name a few, $t$ can be a logit, probit or complementary log-log link functions \citep{mccullagh1984generalized}. 
\subsubsection{Classification method}
\label{classification}
For a given link, the inferential procedure will be divided into two steps. First, we  compute the distribution of the latent process on the test data \begin{equation}
p(f_*|\mathbf{X, y, X_*})=\int
p(f_*|\mathbf{X, f, X_*})p(\mathbf{f}|\mathbf{X, y}) d\mathbf{f},
\label{5}
\end{equation}
where $p(\mathbf{f}|\mathbf{X, y})=p(\mathbf{y}|\mathbf{f})p(\mathbf{f}|\mathbf{X})/p(\mathbf{y}|\mathbf{X})$. Then, we estimate the conditional 
probability of $y_* = 1$ by 
\begin{equation}
p(y_*=1 \ \large | \ \mathbf{X, y, X_*})=\int t(f_*)p(f_*|\mathbf{X, y, X_*})d{f_*}, 
\label{7}
\end{equation}
which is approximately a weighted average of the probability of $y_* = 1$ over all possible realizations of predicted stochastic components that is a Gaussian process.
 
It should be pointed out that both of the two integrands in Equations (\ref{5}) and (\ref{7}) do not have closed forms. For Equation (\ref{7}), following the argument in \citet{williams2006gaussian}, numerical tools such as Monte Carlo method can be used to obtain the approximate value of the integral given $p(f_*|\mathbf{X, y, X_*})$. To obtain Equation~(\ref{5}), \citet{williams1998bayesian} introduced Laplace approximation for this problem. \citet{minka2001family} proposed an alternative expectation propagation(EP). Besides these methods, a number of MCMC algorithms have also been considered. In the following section, we will follow the direct Laplace approximation. 

From Equation (\ref{5}), we can write the approximate distribution of $p(\mathbf{f}|\mathbf{X, y})$ as $N(\hat{\mathbf{f}}, \hat{I}^{-1}),$ where 
$\hat{\mathbf{f}}$ is the MLE of the distribution and $\hat{I}$ is the observed Fisher information matrix.
% We have that \begin{align}
%\log p(\mathbf{f}|\mathbf{X, y})&=\log p(\mathbf{y}|\mathbf{f})+\log p(\mathbf{f}|\mathbf{x})+C\\
%&=\log p(\mathbf{y}|\mathbf{f})-\frac{1}{2} \mathbf{f^T}K^{-1}\mathbf{f}-\frac{1}{2}\log |K|+C
%\end{align}
%Then, it follows that
%\begin{align}
%\nabla\log p(\mathbf{f}|\mathbf{X, y})&=\nabla \log p(\mathbf{y}|\mathbf{f})-K^{-1}\mathbf{f}\\
%\nabla\nabla\log p(\mathbf{f}|\mathbf{X, y})&=\nabla \nabla \log p(\mathbf{y}|\mathbf{f})-K^{-1}
%\end{align}
To find the value of $\hat{\mathbf{f}}$, Newton's method can be implemented, where in each iteration 
$\mathbf{f^{\text{new}}}=\mathbf{f^{\text{old}}}-\nabla^2\log p(\mathbf{f^{\text{old}}}|\mathbf{X, y})^{-1}\nabla\log p(\mathbf{f^{\text{old}}}|\mathbf{X, y})=(K^{-1}(\mathbf{X}, \mathbf{X})+W)^{-1}(W\mathbf{f^{\text{old}}}+\nabla \log p(\mathbf{y}|\mathbf{f^{\text{old}}}))$, where $W=- \nabla^2 \log p(\mathbf{y}|\mathbf{f^{\text{old}}})$ and $K (\mathbf{X, X})$ is the covariance matrix of $f(\mathbf{X})$. Thus, the distribution $p(\mathbf{f}|\mathbf{X, y})$ can be approximated by $N(\hat{\mathbf{f}}, (K^{-1}(\mathbf{X, X})+W)^{-1}).$

\citet{opper1999gaussian} suggested the conditional expectation of $f_*$ could be obtained by 
$\mathbb{E}(f_*|\mathbf{X, y, X_*})=K\mathbf{(X_*, X)}^TK^{-1}(\mathbf{X, X})\hat{\mathbf{f}}=K\mathbf{(X_*, X)}^T\nabla \log p(\mathbf{y}|\hat{\mathbf{f}}).$
Following similar arguments, the conditional variance of  $f_*$ can be obtained by
$\mathbb{V}(f_*|\mathbf{X, y, x_*})
%={K}(\mathbf{X_*},\mathbf{X_*})-\mathbf{k_*}^TK^{-1}\mathbf{k_*}+\mathbf{k_*}^TK^{-1}(K^{-1}+W)^{-1}K^{-1}\mathbf{k_*} %
={K}(\mathbf{X_*},\mathbf{X_*})-K(\mathbf{X_*, X})^T(K^{-1}(\mathbf{X, X})+W)^{-1}K(\mathbf{X_*, X}).$
Given the mean and variance, at the last step, the probability of $y_* = 1$ can be approximated by $\int t(f_*)\hat{p}(f_*|\mathbf{X, y, X_*})d{f_*}.$
It should be pointed out that the Gaussian process essentially captures information beyond those provided by past value of both endogenous and exogenous time series.  
\begin{remark}
$\frac{\partial ^2}{\partial \mathbf{f_i}^2}\log p(y_i|\mathbf{f_i})$ takes the following forms for the logit and probit links, respectively,
\begin{eqnarray*}
\frac{\partial ^2}{\partial \mathbf{f_i}^2}\log p(y_i|\mathbf{f_i}) &=& -p(y_i = 1|\mathbf{f_i})p(y_i = 0|\mathbf{f_i})\\
\frac{\partial ^2}{\partial \mathbf{f_i}^2}\log p(y_i|\mathbf{f_i}) &=& -\frac{\phi(\mathbf{f_i})^2}{\Phi((2y_i - 1)\mathbf{f_i})^2} - \frac{(2y_i - 1)\mathbf{f_i}\phi(\mathbf{f_i})}{\Phi((2y_i - 1)\mathbf{f_i})}
\end{eqnarray*}
Here $\phi(.)$ and $\Phi(.)$ are the normal probability density 
function and the cumulative distribution function, respectively.
\end{remark}

\section{HIBITS: The hybrid estimation method for modeling and predicting binary time series}
\label{twostage}
Building on the established theoretical foundations of Gaussian processes, we now develop a novel two-stage inference and classification method. This section is organized as follows: 
in Section~\ref{mainmotiv}, we discuss the motivation of using the hybrid strategy in modeling sleep stage;  followed by details of the two-stage hybrid method in 
Section~\ref{algo}; 
%methodology of the proposed approach. 
in Section~\ref{model selection}, we discuss our model selection strategy; and in 
Section~\ref{infer}, we provide a method in providing point and interval estimates 
of the coefficients of the covariates and endogenous variables. 

\subsection{Motivation}
\label{mainmotiv}
The common approach is to use a Gaussian distribution with zero mean value as a random effect if the latent process yields, equally likely, positive and negative fluctuations around 0 \citep{kuss2006gaussian}. Yet, when it comes to real data, this set up overlooks the linear structure between covariates and the actual response of interest. For instance, to model the binary sleep state, scientists believe that body temperature and heart rate should be involved as potential predictors. In  \cite{fokianos2002regression}, a regression-based approach for modeling covariates 
is proposed. However, if we naively utilize the existing Gaussian distribution with zero mean function to model the data, the latent process equally produces positive 
and negative value fluctuating around 0 which can 
produce misleading results because it will render the effects of covariates (body temperature and heart rate) to be insignificant. In addition, incorporating those covariates in the covariance function is a reasonable approach to modeling the association. 
However, the interpretation is complicated. Much work has been done to overcome the aforementioned limitations. To name a few, \citet{snelson2004warped} proposed an approach to transform data in agreement with the Gaussian process model. Their work generalized the Gaussian process by warping the observational space. Although the transformed data can be fitted by Gaussian process, it leads to difficulty in the interpretation of the transform. Another drawback is that the effects of particular covariates 
could be lost (or difficult to interpret). \citet{cornford1998non} suggested a Gaussian process regression model with mean function $m(x)=\mathbf{\beta^T x}$. Their work incorporates the effect of particular covariates. The main drawback is the computational burden that results from the choice of hyperparameter and MCMC sampler when it applies to classification problem. Building on the prior work, we develop a two-stage method that takes advantage of the strengths of the existing methods. It is able to model the linear association with particular covariates while maintaining computational efficiency. 

\subsection{The proposed HIBITS  method}
\label{algo}
Consider the data $\{(\bm{x_{1,i}, x_{2,i}},y_i)\}$ where $y_i\in \{1, 0\}, \mathbf{x_{1,i} \in R^p}, \mathbf{x_{2,i} \in R^q}.$ Here, $\mathbf{x_{1, i}}$ are the covariates in the fixed effects part and $\mathbf{x_{2, i}}$ are covariates in the stochastic part. Then, $\prob(y_i=1|\mathbf{x_{1,i}, x_{2,i}})=t(\eta(\mathbf{x_{1,i}, x_{2,i}})).$ We now propose the systematic component 
of the generalized linear model to take the form 
%\begin{equation*}
\[
\eta(\mathbf{x_{1,i}, x_{2,i}})=\mathbf{\beta^Tx_{1,i}}+\mathbf{f(x_{2,i})}
%\end{equation*}
\]
where $\mathbf{\beta^T \in R^p}$ and $\mathbf{f(x_{2,i})} \sim \mathcal{GP}(\mathbf{0}, K(\mathbf{x_{2,i}})).$ The systematic component with fixed and random effects follow a linear mixed effect model with the first part capturing the fixed effect and the second part describing the randomness that is not covered by the first part. Note that $\eta(\mathbf{x_{1,i}, x_{2,i}})$ does not include an intercept term on this stage. Following the same notation as previous sections, we denote $\mathbf{X}_d = (\mathbf{x}_{d,1}, \cdots, \mathbf{x}_{d, n}), {d} = 1, 2$ as the training dataset and $\mathbf{X}_{d*} = (\mathbf{x}_{d,n+1}, \cdots, \mathbf{x}_{d, n+n_*}), {d} = 1, 2$ as the testing subsets. The proposed inference method proceeds as follows. 

%$\bm{\beta}, \beta$
%$\bullet$ 

\vspace{0.25in} 

\noindent \textbf{Stage 1.} Inference on the fixed effect. 

The joint likelihood function $L(\beta|\mathbf{X_1, X_2, y, f(X_2)})$ can be written as 
\begin{equation}L(\beta|\mathbf{X_1, X_2, y, f(X_2)})=\prod_{i=1}^{n}t(\eta(\mathbf{x_{1,i}, x_{2,i}}))^{y_i}(1-t(\eta(\mathbf{x_{1,i}, x_{2,i}})))^{1-y_i}.
\label{like}
\end{equation}
On the first stage, we consider the latent Gaussian process $\mathbf{f(X_2)}$ fixed across time $i$. 
Numerical algorithms such as Newton-Raphson method can be used to obtain $\mathbf{\hat{\beta}}$, the MLE  of the joint likelihood function. In fact, in this stage, we regard the latent Gaussian process $\mathbf{f(X_2)}$ as the time-invariant intercept of the logistic regression, which is considered fixed but unknown. 

%$\bullet$ 

\vspace{0.25in}

\noindent \textbf{Stage 2.} Inference on the stochastic components. 

On the second stage, we make use of the result of inference on the 
fixed effect from Stage 1 and adjust the estimates by introducing the latent Gaussian process $\mathbf{f(X_2)}$.
Conditional on $\mathbf{\hat{\beta}},$ we define $\mathbf{\tilde{\eta}(\mathbf{x_{1,i}, x_{2,i}|\hat{\beta}})}=\mathbf{\hat{\beta}^T x_{1,i}},$ then it follows that $$\prob(y_i=1|\mathbf{x_{1,i}, x_{2,i}},\mathbf{\hat{\beta}})=t(\mathbf{\tilde{\eta}(\mathbf{x_{1,i}, x_{2,i}|\hat{\beta}})}+\mathbf{f(x_{2i})}).$$
Here, we model the stochastic component $\mathbf{f(X_2)}$ as a Gaussian process with covariance function 
 \begin{equation}Cov(\mathbf{f(x_{2,i})}, \mathbf{f(x_{2,j})})=\lambda\exp(-\rho||\mathbf{x_{2,i}}-\mathbf{x_{2,j}}||^2)+\sigma^2\delta_{ij} \label{covariance}\end{equation}
and $\delta_{ij}$ takes value 1 when $i=j$ and 0 otherwise. The parameters $\rho, \sigma$ and $\lambda$ are estimated by the strategy proposed by Section~\ref{model selection} and we will not specify any prior on those parameters. 
Since $\mathbf{\tilde{\eta}(\mathbf{x_{1,i}, x_{2,i}|\hat{\beta}})}$ is known, we can implement the strategy in Section~\ref{gpic} in dealing with the predictive probability from Equation (\ref{7}).
The complete hybrid method can be summarized in the following Algorithm~\ref{al1}. 
\begin{algorithm}

\caption{The proposed binary hybrid method}
\label{al1}
\noindent \textbf{Stage 1.}\\
\textbf{Input}: $\mathbf{y}$, ${K(\mathbf{X_2, X_2})}$(covariance matrix), $p(\mathbf{y}|\mathbf{X_1, f})$ 
(the likelihood function)\\
%\textbf{Procedure}: \\
\textbf{Compute} the MLE $\hat{\mathbf{\beta}}$ of $L(\beta|\mathbf{X_1, X_2, y, f(X_2)})$ using Newton-Raphson method (see Equation~(\ref{like})).\\
$\mathbf{f:=0}$ \quad initialization\\
$\mathbf{While}$ (iter $<$ Max-iter)\\
$\mathbf{Repeat}\\$
$W:=-\nabla^2 \log p(\mathbf{y|\hat{\mathbf{\beta}},f})$\\
$C:=W*\mathbf{f}+ \nabla \log p(\mathbf{y|\hat{\mathbf{\beta}},f})$\\
$\mathbf{f}=(K^{-1}(\mathbf{X_2, X_2})+W)^{-1}*C $\\
$\mathbf{If}$ the difference of successive value of $\mathbf{f}$ is small enough, $\mathbf{break}$\\
$\mathbf{else}$ continue this procedure.\\
$\mathbf{Return}$: $\hat{\mathbf{f}}:=\mathbf{f}$\\
%\end{algorithm}
%\begin{algorithm}

\noindent \textbf{Stage 2.} \\ 
\textbf{Input}: $\mathbf{y}$, $\hat{\mathbf{\beta}}$ (the estimates of coefficients of the fixed effect), $\hat{\mathbf{f}}$ (the mean of the Laplace approximation), ${K}(\mathbf{X_2, X_2}), K(\mathbf{X_{2*}, X_2}), K(\mathbf{X_{2*}, X_{2*}})$(covariance matrix), $p(\mathbf{y}|\mathbf{X_1, f})$(the likelihood function), $\mathbf{X_{1*}, X_{2*}}$ (test input)\\
%\textbf{Procedure}: \\
$W:=-\nabla^2 \log p(\mathbf{y|\hat{\mathbf{\beta}},\hat{f}})$\\
$\mathbf{\bar{f}}_*=K(\mathbf{X_{2*}, X_2})^T \nabla \log p(\mathbf{y|\hat{\mathbf{\beta}},\hat{f}})$\\
$\mathbf{v_*}=K(\mathbf{X_{2*},X_{2*}})-K(\mathbf{X_{2*}, X_2})^T W^{\frac{1}{2}}(I+W^{\frac{1}{2}}{K}(\mathbf{X_2, X_2})W^{\frac{1}{2}})^{-1}W^{\frac{1}{2}}K(\mathbf{X_{2*}, X_2})$\\
$\bar{\pi}_*=\int t(\mathbf{\hat{\beta}}^T\mathbf{X_{1*}}+z) N(z|\mathbf{\bar{f_*}}, \mathbf{v_*})dz$\\
$\mathbf{Return}$: $\bar{\pi}_*$ (the predictive probability of test input $\mathbf{X_{1*}, X_{2*}}$)\\
In the implementation of this method, we conducted a model selection strategy on the covariance matrix ${K}$ based on maximum likelihood in Equation (\ref{mlh}).
\end{algorithm}
\begin{remark}
The Hessian matrix $W$ is a diagonal matrix with the following elements for the logit and probit link respectively,
\begin{eqnarray*}
W_{ii} &=& -p(y_i = 1|\hat{\beta}, \mathbf{f_i})p(y_i = 0|\hat{\beta}, \mathbf{f_i}),\\
W_{ii} &=&-\frac{\phi^2((2y_i - 1)(\mathbf{\mathbf{\hat{\beta}^Tx_{1,i}}+f_i}))(\mathbf{\mathbf{\hat{\beta}^Tx_{1,i}}+f_i})}{\Phi^2((2y_i - 1)(\mathbf{\mathbf{\hat{\beta}^Tx_{1,i}}+f_i}))} - \frac{(2y_i-1)(\mathbf{\mathbf{\hat{\beta}^Tx_{1,i}}+f_i})\phi(y_i(\mathbf{\mathbf{\hat{\beta}^Tx_{1,i}}+f_i}))}{\Phi((2y_i-1)(\mathbf{\mathbf{\hat{\beta}^Tx_{1,i}}+f_i}))}.
\end{eqnarray*}
\end{remark}
%\newpage
\subsection{Model selection}\label{model selection}
Strategies on model selection are also presented in two steps. 
%$\bullet$ 

\noindent \textbf{Step 1. } In this study, we will use exploratory analysis to choose variables. Alternatively, we could use AIC or BIC focusing on the fixed effects. Using automatic variable selection strategies based on AIC or BIC, we can choose a model with a subset of predictors. AIC value is defined as $\text{AIC} = 2k - 2 \log L$ and BIC is defined as $\text{BIC} = k \log n - 2 \log L$, where $k$ is the number of parameters, $n$ is the number of observations and $L$ is the maximum value of likelihood.  

%$\bullet$ 
\vspace{0.25in}

\noindent \textbf{Step 2.} We select the parameters for the covariance matrix by maximum likelihood estimation. The strategy is inspired by the work of \citet{williams2006gaussian}. Our work is similar in terms of maximizing the 
marginal likelihood but differs in the way that the both fixed and random effects are involved. 
	
We denote $\theta$ as the parameters in the covariance structure $Cov (\mathbf{y})$. The approximate log marginal likelihood is 
	\begin{equation}
	\log q(\mathbf{y}|\mathbf{X_1, X_2},\theta)=-\frac{1}{2}\hat{\mathbf{f}}^TK^{-1}(\mathbf{X_1, X_1})\hat{\mathbf{f}}+\log p(\mathbf{y}|\mathbf{X_1}, \hat{\mathbf{f}})-\frac{1}{2}\log |B| ,
\label{mlh}
\end{equation}
where $B=I+W^{\frac{1}{2}}K(\mathbf{X_1, X_1})W^{\frac{1}{2}}$ and $\hat{\mathbf{f}}$ is defined in Section~\ref{classification}. The strategy is to choose the value of $\theta$ that maximizes Equation (\ref{mlh}). Note that the covariance matrix $K$ ($K(\mathbf{X_1, X_1})$)and $\hat{\mathbf{f}}$ involve parameters $\theta$, the partial derivative of $\frac{\partial \log q(\mathbf{y}|\mathbf{X_1, X_2},\theta)}{\partial \theta_j}$ is therefore 
	$$
	\frac{\partial \log q(\mathbf{y}|\mathbf{X_1, X_2},\theta)}{\partial \theta_j}=A+B,
	$$
	where $A$ and $B$ are defined as follows
	\begin{align*}
	A&=\frac{1}{2}\hat{\mathbf{f}}^TK^{-1}\frac{\partial K}{\partial \theta_j}K^{-1}\hat{\mathbf{f}}-\frac{1}{2}tr((W^{-1}+K)^{-1}\frac{\partial K}{\partial \theta_j}), \\
	B&=\sum_{i=1}^{n}-\frac{1}{2}[(K^{-1}+W)^{-1}]_{ii}\frac{\partial^3}{\partial f^3_i}\log p(\mathbf{y}|\mathbf{X_1}, \hat{\mathbf{f}})[(I+KW)^{-1}\frac{\partial K}{\partial \theta_j}\nabla\log p(\mathbf{y}|\mathbf{X_1}, \hat{\mathbf{f}})]_i.
	\end{align*}
Newton-Raphson method or coordinate descent will be applied to optimize the log marginal likelihood in Equation (\ref{mlh}).

In this study, the parameters $\theta$ from Equations (\ref{covariance}) are $\rho, \sigma$ and $\lambda.$ Through our simulation studies, we specify the parameters $\sigma$ and $\rho$ and apply the aforementioned strategy on estimating $\lambda$ for the following reasons: (1) it might lead to identifiability problem if we do not fix some of the parameters in this frequentist setting; (2) results do not show much difference if parameters $\sigma$ and $\rho$ are not fixed; (3) computation will be demanding if no parameter is fixed. 

\subsection{Inference on the effects of covariates}
\label{infer}
We propose to use bootstrap sampler to provide point and confidence intervals of the linear coefficients of the covariates $\mathbf{X_1}$. Our approach is based on resampling the stochastic component and maximum likelihood. The algorithm is summarized in Algorithm~\ref{a2}.
\begin{algorithm*}
	\caption{Inference on the linear effects}
	\label{a2}
	\textbf{Input}: $\mathbf{y}$, ${\hat{K}}$(the estimated covariance matrix derived from Section~\ref{model selection}), $\hat{\mathbf{\beta}}$ (the estimates of coefficients of the fixed effect derived from Stage 1 in Section~\ref{algo})\\
	\textbf{Procedure}: \\
	$\tilde{\eta} (\mathbf{X_1}):= \hat{\beta}^T\mathbf{X_1}$\\
	\textbf{While} (Iter $<$ Max-iter)\\
	\textbf{Repeat}\\
	\textbf{Generate} $\mathbf{f^{\textbf{(iter)}}(X_2)}$ from $\mathcal{GP}$ with covariance function $ {\hat{K}}$\\
	${\eta}^\textbf{(iter)} (\mathbf{X_1, X_2}):= \tilde{\eta} (\mathbf{X_1})+ \mathbf{f^{\textbf{(iter)}}(X_2)}$\\
	\textbf{Compute} the MLE $\hat{{\beta}}^{(\textbf{iter})}$ of $L(\beta|\mathbf{X_1, X_2, y, f(X_2)})$ using Newton-Raphson method, where 
	$
	L(\beta|\mathbf{X_1, X_2, y, f(X_2)})=\prod_{i=1}^{n}t({\eta}^{(\textbf{iter})}(\mathbf{x_{1,i}, x_{2,i}}))^{y_i}(1-t(\eta^{(\textbf{iter})}(\mathbf{x_{1,i}, x_{2,i}})))^{1-y_i}
	$\\
	\textbf{End of while}\\
	\textbf{Compute} ${\hat{\beta_*}} = \frac{1}{\text{Max-iter}}\sum_{i = 1}^{\text{Max-iter}}\hat{{\beta}}^{(i)}$\\
	$\hat{\beta}_{0.025} = 2.5$-th percentile of $\{\hat{\beta}^{(i)}\}_{i=1}^{\text{Max-iter}}$\\
		$\hat{\beta}_{0.975} = 97.5$-th percentile of $\{\hat{\beta}^{(i)}\}_{i=1}^{\text{Max-iter}}$\\
  \textbf{Return:}  ${\hat{\beta_*}}$ (The point estimates of the parameters from linear effects); $(\hat{\beta}_{0.025}, \hat{\beta}_{0.975})$ (The $95 \%$ bootstrap confidence interval of the parameters from linear effects).
\end{algorithm*}
\subsection{Summary}
In summary, the proposed method on inference, prediction and model selection 
maintain the following strengths: (1.) it uses linear and non-linear stochastic 
components to model the effect of the covariates on the response; (2.) it provides point and 
interval estimates of the linear effects that are more efficient than the 
existing methods as demonstrated in Section~\ref{simulations}; (3.) it is able to 
make accurate predictions as shown in Section~\ref{simulations}; (4.) the 
computational cost is not demanding; (5.) similarly to generalized linear models, 
it provides results that are straightforward to interpret.
 
\section{Simulations}
\label{simulations}
In this section, simulations are implemented to test the performance of the 
proposed method. In Section~\ref{logit}, binary time series $\mathbf{y_i}$ 
are generated by the logit model. We compared the classification error rates 
derived from the proposed method with 6 other competing statistical and machine learning approaches, namely, the ordinal model, logistic regression, generalized additive mixed model, random forest, decision tree and gradient boosting. We also compute the point and confidence intervals of the coefficients of the covariates and endogenous variables in comparison with other existing methods. To test the robustness of our method, in Section~\ref{probit}, we generate time series $\mathbf{y_i}$ from the probit model but use the logit model to fit the data.  In Section~\ref{kernel}, we utilize mixture kernels to generate the Gaussian process and then apply the proposed HIBITS method. Classification error rates, point estimates and confidence intervals are also utilized as measures for comparison.
 
\subsection{Prediction and inference performance on logit model}
 \label{logit}
\begin{figure}[th!] \centering
                            \begin{tabular}{cc}
                            \includegraphics[width=.5\textwidth]{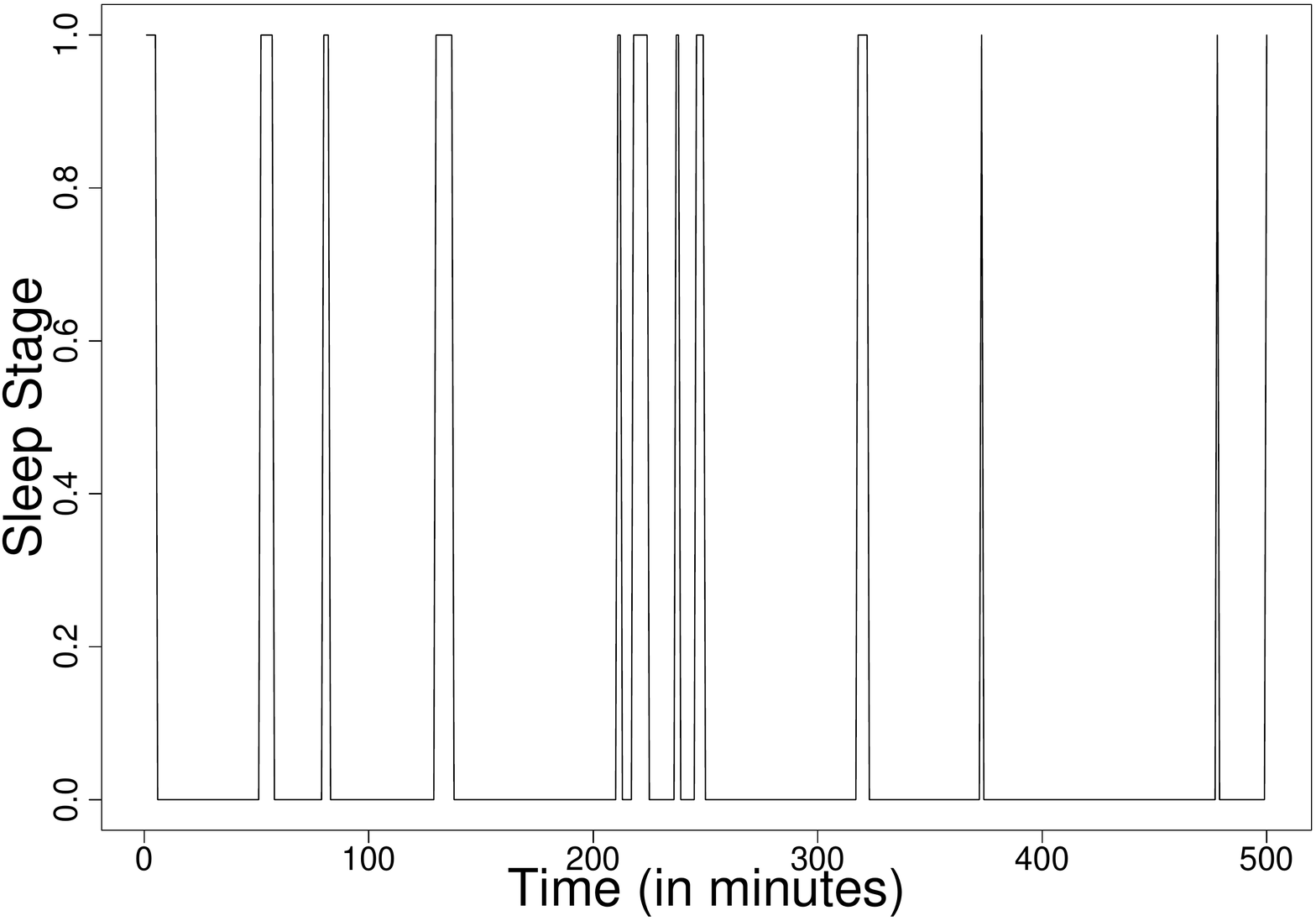} &
                            \includegraphics[width=.5\textwidth]{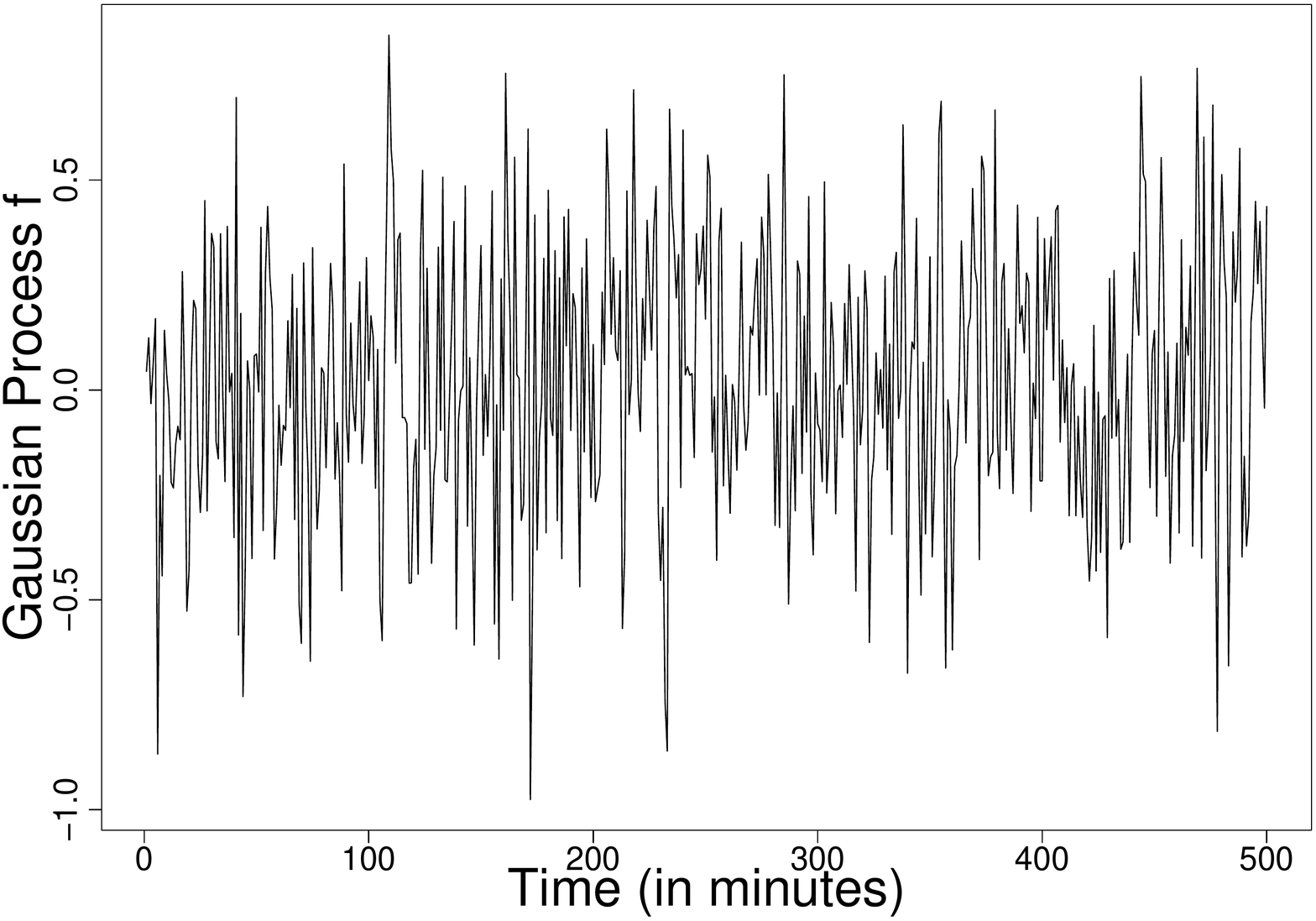} \\
                            \end{tabular}
                            \caption{Plots of the generated sleep stage (left) and the simulated Gaussian process(right).}
                            \label{sim}
                            \end{figure}

To evaluate the prediction power and robustness of the proposed method, 
binary time series ${y_i}$ are generated under two scenarios: 
\begin{itemize}

\item {\bf Scenario 1 (with a stochastic process).}\\
 $\prob ({y_i}=1)=logit^{-1}(\beta_0{x_{1i}}+\beta_1{y_{i-1}}+\mathbf{f}(x_{2i}))$; 
\item {\bf Scenario 2 (without a stochastic process).}\\
$\prob ({y_i}=1)=logit^{-1}(\beta_0{x_{1i}}+\beta_1{y_{i-1}})$. \\
\end{itemize}
Here, $\mathbf{f(x_2)}$ follows Gaussian process with $$Cov(\mathbf{f}(x_{2i}), \mathbf{f}(x_{2j}))=\lambda\exp(-\rho({x_{2i}}-{x_{2j}})^2)+\sigma^2\delta_{ij}$$
and $\delta_{ij}$ takes value 1 when $i=j$ and 0 otherwise. The parameter $\beta_1$ controls the strength of dependence on previous realizations ${y_{i-1}}$ and it denotes the log odds ratios of ${y_{i-1}}=1$ versus ${y_{i-1}}=0$. $\beta_0$ is the linear coefficients with respect to covariates at current time point. $\lambda$ is the parameter that determines the strength of dependence across adjacent time points. In this simulation, parameters $\mathbf{\beta}= (\beta_0, \beta_1)$ and $\lambda$ vary in different scenarios. $1000$ simulations are conducted in each scenario. 
Figure \ref{sim} shows plots of the simulated data. In this scenario, $\mathbf{\beta}=(0.5, 4), \lambda=1, \rho=1, \sigma=0.1.$ 500 sleep stages were generated.  

\noindent {\bf Alternative Methods.} To evaluate the prediction power of the proposed method, we compare the classification error rates with other six competing approaches. In general, those approaches include regression and tree based classification strategies. Generalized linear model with logit link is fitted as the first competing method. Further,  to respect the correlated structure of the binary time series, we consider the generalized additive mixed models (GAMMs) as the second regression based competing approach. In the work of \citet{lin1999inference}, linear structures of covariates are extended to be smooth functions. Following the notation in Section~\ref{twostage}, the GAMM model is defined as 
$$
\eta({\mathbf{x_{1,i}}})=\beta_0 + f_1({x_{1,i}^1}) + \cdots + f_p({x_{1,i}^p}) + \mathbf{z_i^Tb_i},
$$                            
 where ${x_{1,i}^j}$ denotes the $j^{th}$ component of vector $\mathbf{x_{1,i}}$, $f_j(.)$ is a centered twice-differentiable smooth function, the random effects $\mathbf{b}$ are assumed to be distributed as $N(0, D(\mathbf{\theta}))$ and $\theta$ is the variance components. \citet{lin1999inference} estimated nonparametric functions and parameters by using smoothing splines and marginal quasi-likelihood. In this simulation, R package `gamm4' was implemented to test the performance of GAMMs. We also considered the regression models for nominal and ordinal time series introduced by \citet{fokianos2002regression}. As is discussed in their concrete work, we implemented ordinal time series model in the simulation. It should be pointed out that due to the binary response, ordinal time series model is degenerated into logistic regression. Simulation results also suggest the equivalence of these two approaches. In addition, we compared our method to tree-based classification approaches. In general, we split the feature space (heart rate and previous sleep states in this study) into ``subspaces" and fit simple models within each region. Following the derivation in \citet{friedman2001elements}, for a node $m$ denoting a region $R_m$ with $N_m$ observations, we let 
 $$
 \hat{p}_{mk}=\frac{1}{N_m}\sum_{\mathbf{x_{1,i}}\in R_m}\mathbf{1}({y_i}=k),
 $$
 where class $k$ is either $0$ or $1$ and $\mathbf{1}$ is the indicator function. We assign the observations in node $m$ to class $k(m)=\arg \max_k\hat{p}_{mk}.$ Measures of node impurity, denoted as $Q_m(T)$,  can be chosen as 
 the misclassification error, Gini index and cross-entropy or deviance. 
 
To further extend the decision tree approach, we also consider random forest and gradient boosting algorithms in the simulation. The essence of random forest is to average many noisy but asymptotically unbiased classifiers and hence reduce the variation. It requires bootstrapping samples and selection features from the training dataset. Since there exist only a few features in this model, the benefit from using random forest approach is mainly derived from the bootstrapping strategy.  For each bootstrap training sample set, we grow a random forest tree $T_b, b=1, \cdots, B.$ The final output is the ensemble of trees and then predictions are made 
by majority vote. 
 In addition to random forest, gradient boosting is another extension of decision tree based method. Similar to the general boosting methods, gradient boosting searches for a strategy to combine multiple weak classifiers in an iterative manner. As discussed in \citet{friedman2001greedy} and \citet{friedman2001elements}, the method generically starts from a model with constant value. At iteration $m (1 \leq m \leq M)$, suppose the classifier is denoted as $F_{m-1}(\mathbf{x_1, x_2})$, we calculate pseudo-residuals by 
 $$
 r_{im}=-\Big[\frac{\partial L({y_i}, F(\mathbf{x_{1,i}, x_{2, i}}))}{\partial F(\mathbf{x_{1, i}, x_{2, i}}))}\Big]_{F(\mathbf{x_{1, i}, x_{2, i}})=F_{m-1}(\mathbf{x_{1, i}, x_{2, i}})},
 $$where $L(y, F(x))$ is a loss function. Then, we fit a classifier $h_m(x)$ to the pseudo-residuals and implement a line search algorithm in solving the optimization problem 
 $$
 \gamma_m=\arg\min_\gamma\sum_{i=1}^{n}L(y_i, F_{m-1}(\mathbf{x_{1, i}, x_{2, i}})+\gamma h_m(\mathbf{x_{1, i}, x_{2, i}})).
 $$
 At the end of this iteration, we update the model by 
 $$
F_{m}(\mathbf{x_{1, i}, x_{2, i}}) = F_{m-1}(\mathbf{x_{1, i}, x_{2, i}})+\gamma_m h_m(\mathbf{x_{1, i}, x_{2, i}}).
$$
% \mathbf{x_{2,i}}
We keep repeating the full sweep until convergence. The final classifier is denoted as $F_{M}(\mathbf{x_{1, i}, x_{2, i}})$.\\
\vspace{0.25in}\\
{\bf Model Evaluation.} To formally evaluate the performance of all the aforementioned approaches, we calculate the  classification error rates under both scenarios. In particular, we fit the results in linear mixed effect model to account for the correlation among classification errors across different methods that result from the same simulated dataset. We consider the model  
$$
E_{ij} = \mu_i + z_j + \epsilon_{ij}, 
$$
where $E_{ij}$ denotes the classification error rate of approach $i$ on dataset $j$; $\mu_i$ is the mean error rate of method $i$, which is well-defined by the law of large numbers. $z_j  \overset{iid}{\sim} N ( 0, \sigma^2),$ $\epsilon_{ij} \overset{iid}{\sim} N ( 0, \tau^2),$ $i = 1, \cdots, 6$ and $j = 1, \cdots, 1000.$  We calculate the simultaneuous $95 \%$ Bonferroni  confidence intervals of $(\mu_1 - \mu_i), i = 2, \cdots, 6$ to detect the difference in the mean error rates between the proposed method with all the other approaches. In particular, $\mu_1, \cdots, \mu_6$ denote the mean error rates of HIBITS, Ordinal model (logistic regression), GAMMs, Random forest, Gradient boosting and Decision tree respectively.

Table~\ref{sim4_1} provides a summary of the simulation studies for various parameters. It can be seen that for datasets with Gaussian process, there is statistically significant difference in comparison with the competing methods. In particular, the proposed HIBITS method produces significantly lower prediction error rates compared to existing methods. The advantage of the proposed approach is even more obvious when compared with gradient boosting and decision tree approaches. The results show that the proposed HIBITS method captures effective information from covariates ${x_{1i}}$, $\mathbf{y_{i-1}}$ and also the stochastic process. The covariate $\mathbf{y_{i-1}}$ serves as a significant predictor as we increase the ratio of $\beta_1$ over $\beta_0$. 

For the datasets generated {\it without} the Gaussian process (Scenario 2) shown in Table~\ref{sim4_2}, the accuracy prediction from the two-stage approach is significantly higher than some of the existing approaches such as decision tree and gradient boosting. Among all the other competitors, the proposed method behaves equally competitive. This shows the robustness of the proposed approach when data have no Gaussian process components. This is partly due to the strategy on choosing hyperparemeters. By controlling their values, the effects of Gaussian process will be adjusted to the data. 
\begin{table}[H]

\caption{Summary of simulation results. $\mu_1, \cdots, \mu_6$ denote the mean error rates of HIBITS, Ordinal model (logistic regression), GAMMs, Random forest, Gradient boosting and Decision tree respectively. $1000$ simulated datasets were generated under the scenario: $\prob ({y_i}=1)=logit^{-1}(\beta_0{x_{1i}}+\beta_1{y_{i-1}}+\mathbf{f}(x_{2i}))$ (``Scenario 1''). We calculated the $95\%$ Bonferroni-corrected confidence intervals of the prediction error difference  from the testing dataset, $\mu_1 - \mu_i, i=2, \cdots, 6$ that the classification error rate for the proposed method is lower than that for each of competing methods.}

	\centering
\begin{small}
\begin{tabular}{lllccc}
\hline
&&& \multicolumn{2}{c}{Scenario 1}& \\
\cline{4-6} 
\multicolumn{2}{c}{Parameters($\beta, \lambda$)}& Competing Method & \multicolumn{2}{c}{95\% confidence interval of } &  \\
\multicolumn{2}{c}{} & &  \multicolumn{2}{c}{$\mu_1 - \mu_i, i=2, \cdots, 7$} &\\
\hline

$\beta=(0.5,3), \lambda=10$ & & Ordinal model* & \multicolumn{2}{c}{${(-0.052 \quad , \quad -0.032)} $  }  & \\
&     & GAMMs&       \multicolumn{2}{c}{${(-0.052\quad , \quad -0.032)} $  } &   \\
&     & Random forest &        \multicolumn{2}{c}{${(-0.029 \quad , \quad -0.009)} $  }   &  \\
%&     & Ordinal model&        \multicolumn{2}{c}{${(-0.032 \quad , \quad -0.001)} $  } && \multicolumn{2}{c}{${(-0.006 \quad , \quad +0.003)} $  }  \\
&     & Gradient boosting&       \multicolumn{2}{c}{${(-0.075 \quad , \quad -0.055)} $  }&  \\
&     & Decision tree&        \multicolumn{2}{c}{${(-0.070 \quad , \quad -0.050)} $  }  & \\
\hline

$\beta=(0.5,3), \lambda=5$ & & Ordinal model & \multicolumn{2}{c}{${(-0.015 \quad , \quad -0.001)} $  }  &  \\
&     & GAMMs&       \multicolumn{2}{c}{${(-0.017\quad , \quad -0.001)} $  } &  \\
&     & Random forest &        \multicolumn{2}{c}{${(-0.017 \quad , \quad -0.002)} $  }   & \\
%&     & Ordinal model&        \multicolumn{2}{c}{${(-0.033 \quad , \quad -0.001)} $  } && \multicolumn{2}{c}{${(-0.025 \quad , \quad +0.007)} $  }  \\
&     & Gradient boosting&       \multicolumn{2}{c}{${(-0.038 \quad , \quad -0.022)} $  }& \\
&     & Decision tree&        \multicolumn{2}{c}{${(-0.048 \quad , \quad -0.032)} $  }  &  \\
\hline

$\beta=(0.5,3.5), \lambda=10$ & & Ordinal model & \multicolumn{2}{c}{${(-0.036 \quad , \quad -0.013)} $  }  &  \\
&     & GAMMs&       \multicolumn{2}{c}{${(-0.030\quad , \quad -0.011)} $  } &  \\
&     & Random forest &        \multicolumn{2}{c}{${(-0.021 \quad , \quad -0.001)} $  }   & \\
%&     & Ordinal model&        \multicolumn{2}{c}{${(-0.033 \quad , \quad -0.001)} $  } && \multicolumn{2}{c}{${(-0.025 \quad , \quad +0.007)} $  }  \\
&     & Gradient boosting&       \multicolumn{2}{c}{${(-0.046 \quad , \quad -0.028)} $  }& \\
&     & Decision tree&        \multicolumn{2}{c}{${(-0.055 \quad , \quad -0.037)} $  }  &  \\
\hline

$\beta=(0.5,3.5), \lambda=5$ & & Ordinal model & \multicolumn{2}{c}{${(-0.010 \quad , \quad -0.001)} $  }  &  \\
&     & GAMMs&       \multicolumn{2}{c}{${(-0.011\quad , \quad -0.001)} $  } &  \\
&     & Random forest &        \multicolumn{2}{c}{${(-0.015 \quad , \quad -0.001)} $  }   & \\
%&     & Ordinal model&        \multicolumn{2}{c}{${(-0.033 \quad , \quad -0.001)} $  } && \multicolumn{2}{c}{${(-0.025 \quad , \quad +0.007)} $  }  \\
&     & Gradient boosting&       \multicolumn{2}{c}{${(-0.020 \quad , \quad -0.006)} $  }& \\
&     & Decision tree&        \multicolumn{2}{c}{${(-0.035 \quad , \quad -0.021)} $  }  &  \\
\hline
\multicolumn{6}{l}{* For binary time series, ordinal model is equivalent to logistic regression.}
\end{tabular}
	\label{sim4_1}
\end{small}
\end{table}

\begin{table}[H]

	\caption{Summary of simulation results. $\mu_1, \cdots, \mu_6$ denote the mean error rates of HIBITS, Ordinal model (logistic regression), GAMMs, Random forest, Gradient boosting and Decision tree respectively. $1000$ simulated datasets were generated under the scenario: $\prob ({y_i}=1)=logit^{-1}(\beta_0{x_{1i}}+\beta_1{y_{i-1}})$ (``Scenario 2''). We calculated the $95\%$ Bonferroni-corrected confidence intervals of the prediction error difference  from the testing dataset, $\mu_1 - \mu_i, i=2, \cdots, 6$ that the classification error rate for the proposed method is lower than that for each of competing methods.}
		\centering
	\begin{small}
	\begin{tabular}{lllccc}
			\hline
			&&& \multicolumn{2}{c}{Scenario 2} &\\
			\cline{4-6} 
			\multicolumn{2}{c}{Parameters($\beta$)}& Competing Method & \multicolumn{2}{c}{95\% confidence interval of }  & \\
			\multicolumn{2}{c}{} & &  \multicolumn{2}{c}{$\mu_1 - \mu_i, i=2, \cdots, 7$} & \\
			\hline
			
			$\beta=(0.5,3)$ & & Ordinal model & \multicolumn{2}{c}{${(-0.006 \quad , \quad +0.010)} $  }  &  \\
			&     & GAMMs&       \multicolumn{2}{c}{${(-0.005\quad , \quad +0.009)} $  } & \\
			&     & Random forest &        \multicolumn{2}{c}{${(-0.004 \quad , \quad +0.012)} $  }   & \\
			%&     & Ordinal model&        \multicolumn{2}{c}{${(-0.032 \quad , \quad -0.001)} $  } && \multicolumn{2}{c}{${(-0.006 \quad , \quad +0.003)} $  }  \\
			&     & Gradient boosting&       \multicolumn{2}{c}{${(-0.022 \quad , \quad -0.001)} $  }& \\
			&     & Decision tree&        \multicolumn{2}{c}{${(-0.023 \quad , \quad -0.002)} $  }  & \\
			\hline
$\beta=(0.5,3.5)$ & & Ordinal model & \multicolumn{2}{c}{${(-0.003 \quad , \quad +0.010)} $  }  &  \\
&     & GAMMs&       \multicolumn{2}{c}{${(-0.002\quad , \quad +0.010)} $  } &  \\
&     & Random forest &        \multicolumn{2}{c}{${(-0.015 \quad , \quad -0.001)} $  }   & \\
%&     & Ordinal model&        \multicolumn{2}{c}{${(-0.033 \quad , \quad -0.001)} $  } && \multicolumn{2}{c}{${(-0.025 \quad , \quad +0.007)} $  }  \\
&     & Gradient boosting&       \multicolumn{2}{c}{${(-0.020 \quad , \quad -0.006)} $  }& \\
&     & Decision tree&        \multicolumn{2}{c}{${(-0.018 \quad , \quad -0.001)} $  }  &  \\
\hline			
\end{tabular}
	\label{sim4_2}
	\end{small}
\end{table}
We also evaluate the performance of modeling the linear effects of covariates ${x_{1i}}$, $\mathbf{y_{i-1}}$ by comparing the $95\%$ confidence intervals of $\beta_0$ and $\beta_1$ with the corresponding interval estimates from the other 
existing methods. Table~\ref{sim4_3} summarizes the results under the same scenarios in Table \ref{sim4_1}. It shows that compared with ordinal model, the proposed HIBITS method produces narrower confidence intervals of parameters $\beta_0$ while 
maintaining high capture rates of the true values. The length difference is obvious and it can gain almost $60\%$ shorter confidence intervals in some scenario. It should be noted that using ordinal model, the capture rate of $\beta_1$ is extremely low while HIBITS method provides promising performance. The same pattern can also be found in Table~\ref{sim4_4}. Under Scenario 2, the benefits of using HIBITS is even more 
obvious in terms of shorter confidence interval length and high capture rate.
\begin{table}[H]
	\caption{Summary of simulation results. $1000$ simulations were generated under the scenario: $\prob ({y_i}=1)=logit^{-1}(\beta_0{x_{1i}}+\beta_1{y_{i-1}}+\mathbf{f}(x_{2i}))$ (``Scenario 1''
		). We present the 95\%  confidence intervals $\beta_0$ and $\beta_1$ from the training dataset.}
	\centering
	\begin{small}
		\begin{tabular}{lllccc}
			\hline
			&&& \multicolumn{2}{c}{Scenario 1} &\\
			\cline{4-6} 
			\multicolumn{2}{c}{Parameters($\beta, \lambda$)}& Method & \multicolumn{2}{c}{95\% confidence interval of }  & \\
			\multicolumn{2}{c}{} & &  $\beta_0$ & $\beta_1$& \\
			\hline
			
			$\beta=(0.5,3), \lambda=10$ & & HIBITS method& $(0.113, 0.547)$&$(1.385, 3.424)$ & \\
			&     & Ordinal model &     $ (-0.292, 0.586)$ & $(0.695, 2.473)$&  \\
		\hline
			
			$\beta=(0.5,3), \lambda=5$ & & HIBITS method& $(0.163, 0.572)$&$(1.570, 3.700)$ & \\
		&     & Ordinal model &     $ (-0.267, 0.637)$ & $(0.850, 2.676)$&  \\
		\hline
			
			$\beta=(0.5,3.5), \lambda=10$ & & HIBITS method& $(0.092, 0.535)$&$(1.628, 4.082)$ &\\
			&     & Ordinal model &     $ (-0.358, 0.625)$ & $(0.806, 2.593)$&   \\
		\hline
		
			$\beta=(0.5,3.5), \lambda=5$ & & HIBITS method& $(0.182, 0.582)$&$(1.820, 3.985)$ & \\
			&     & Ordinal model &     $ (-0.286, 0.694)$ & $(0.991, 2.841)$&  \\

\hline
\end{tabular}
\end{small}
\label{sim4_3}
\end{table}
 \begin{table}[H]
 	\caption{Summary of simulation results. $1000$ simulations were generated under the scenario: $\prob ({y_i}=1)=logit^{-1}(\beta_0{x_{1i}}+\beta_1{y_{i-1}})$ (``Scenario 2''
 		). We present the 95\%  confidence intervals $\beta_0$ and $\beta_1$ from the training dataset.}
 	\centering
 	\begin{small}
 		\begin{tabular}{lllccc}
 			\hline
 			&&& \multicolumn{2}{c}{Scenario 2} &\\
 			\cline{4-6} 
 			\multicolumn{2}{c}{Parameters($\beta$)}& Method & \multicolumn{2}{c}{95\% confidence interval of }  & \\
 			\multicolumn{2}{c}{} & &  $\beta_0$ & $\beta_1$& \\
 			\hline
 			
 			$\beta=(0.5,3)$ & & HIBITS method& $(0.467, 0.600)$&$(2.838, 3.173)$ & \\
 			&     & Ordinal model &     $ (-0.177, 1.420)$ & $(1.677, 4.515)$&  \\
 			\hline

 			$\beta=(0.5,3.5)$ & & HIBITS method& $(0.422, 0.556)$&$(3.468, 3.771)$ &\\
 			&     & Ordinal model &     $ (-0.081, 1.202)$ & $(2.275, 5.096)$&   \\
 			\hline

 		\end{tabular}
 	\end{small}
 \label{sim4_4}
 \end{table}
 
Overall, the proposed method outperforms competing approaches when comparing the results from data both with and without Gaussian process. Through the model selection strategy discussed in Section~\ref{model selection}, the proposed approach can adjust the covariance matrix to the data, which in return produces lower prediction error rate and more efficients inference on covariates than existing methods.

\subsection{Investigating robustness of the estimation method}
\label{probit}
Our goal is to investigate robustness of the proposed model by applying the logistic-based model on data that are generated using a probit model. We generate binary time series $\mathbf{y_i}$ following the scenarios:

\begin{itemize}
\item {\bf Scenario 3 (with a stochastic process).}\\
 $\prob ({y_i}=1)=\Phi(\beta_0{x_{1i}}+\beta_1{y_{i-1}}+\mathbf{f}(x_{2i}))$; 
\item {\bf Scenario 4 (without a stochastic process).}\\
$\prob ({y_i}=1)=\Phi(\beta_0{x_{1i}}+\beta_1{y_{i-1}})$. $\Phi(.)$ is the cumulative distribution function of standard normal distributions and $\mathbf{f(x_2)}$ is defined in the same manner as in Section~\ref{logit}. 
\end{itemize}
Parameters $\beta= (\beta_0, \beta_1)$ and $\lambda$ vary in different scenarios. $1000$ simulations are conducted in each scenario. We fit the same linear mixed effect model discussed in Section~\ref{logit}. Tables~\ref{sim4_5} and~\ref{sim4_6} show the summary of confidence intervals $\mu_1 - \mu_i, i = 2, \cdots, 6.$ Similar to the results in Section~\ref{logit}, for dataset with Gaussian process, most of the confidence intervals do not cover 0. The negative values of the classification error rates imply remarkable benefits of using the proposed method over the other competing methods. Note that when comparing with the gradient boosting and decision tree approaches, the proposed method behaves significantly better in terms of extraordinary higher prediction accuracy. In Scenario 4, we tested the proposed method on the dataset without Gaussian process. It is shown that although there is no significant difference in comparison with other competing methods, the proposed approach produces the same prediction power as other competing methods, which implies the robustness with regard to various link functions. In addition, Table \ref{sim4_7} shows the $95\%$ confidence intervals of the coefficients $\beta_0, \beta_1$ derived from the proposed method and the ordinal model. Similar to the results in Section~\ref{logit}, the proposed method yields much narrower confidence intervals while maintaining good properties of capturing true values. 
\begin{table}[H]
	\caption{Summary of simulation results. $\mu_1, \cdots, \mu_6$ denote the mean error rates of HIBITS, Ordinal model (logistic regression), GAMMs, Random forest, Gradient boosting and Decision tree respectively. $1000$ simulated datasets were generated under the scenario: $\prob ({y_i}=1)=\Phi(\beta_0{x_{1i}}+\beta_1{y_{i-1}}+\mathbf{f}(x_{2i}))$ (``Scenario 3''). We calculated the $95\%$ Bonferroni-corrected confidence intervals of the prediction error difference  from the testing dataset, $\mu_1 - \mu_i, i=2, \cdots, 6$ that the classification error rate for the proposed method is lower than that for each of competing methods.}
	\centering
	\begin{small}
		\begin{tabular}{lllccc}
			\hline
			&&& \multicolumn{2}{c}{Scenario 3}& \\
			\cline{4-6} 
			\multicolumn{2}{c}{Parameters($\beta, \lambda$)}& Competing Method & \multicolumn{2}{c}{95\% confidence interval of } &  \\
			\multicolumn{2}{c}{} & &  \multicolumn{2}{c}{$\mu_1 - \mu_i, i=2, \cdots, 7$} &\\
			\hline
			
			$\beta=(0.5,3), \lambda=10$ & & Ordinal model& \multicolumn{2}{c}{${(-0.042 \quad , \quad -0.023)} $  }  & \\
			&     & GAMMs&       \multicolumn{2}{c}{${(-0.041\quad , \quad -0.022)} $  } &   \\
			&     & Random forest &        \multicolumn{2}{c}{${(-0.025 \quad , \quad -0.006)} $  }   &  \\
			%&     & Ordinal model&        \multicolumn{2}{c}{${(-0.032 \quad , \quad -0.001)} $  } && \multicolumn{2}{c}{${(-0.006 \quad , \quad +0.003)} $  }  \\
			&     & Gradient boosting&       \multicolumn{2}{c}{${(-0.064 \quad , \quad -0.045)} $  }&  \\
			&     & Decision tree&        \multicolumn{2}{c}{${(-0.060 \quad , \quad -0.041)} $  }  & \\
			\hline
			
			$\beta=(0.5,3), \lambda=5$ & & Ordinal model & \multicolumn{2}{c}{${(-0.015 \quad , \quad -0.002)} $  }  &  \\
			&     & GAMMs&       \multicolumn{2}{c}{${(-0.016\quad , \quad -0.002)} $  } &  \\
			&     & Random forest &        \multicolumn{2}{c}{${(-0.021 \quad , \quad -0.007)} $  }   & \\
			%&     & Ordinal model&        \multicolumn{2}{c}{${(-0.033 \quad , \quad -0.001)} $  } && \multicolumn{2}{c}{${(-0.025 \quad , \quad +0.007)} $  }  \\
			&     & Gradient boosting&       \multicolumn{2}{c}{${(-0.024 \quad , \quad -0.009)} $  }& \\
			&     & Decision tree&        \multicolumn{2}{c}{${(-0.040 \quad , \quad -0.026)} $  }  &  \\
			\hline
			
			$\beta=(0.5,3.5), \lambda=10$ & & Ordinal model & \multicolumn{2}{c}{${(-0.030 \quad , \quad -0.001)} $  }  &  \\
			&     & GAMMs&       \multicolumn{2}{c}{${(-0.029\quad , \quad -0.007)} $  } &  \\
			&     & Random forest &        \multicolumn{2}{c}{${(-0.006 \quad , \quad +0.026)} $  }   & \\
			%&     & Ordinal model&        \multicolumn{2}{c}{${(-0.033 \quad , \quad -0.001)} $  } && \multicolumn{2}{c}{${(-0.025 \quad , \quad +0.007)} $  }  \\
			&     & Gradient boosting&       \multicolumn{2}{c}{${(-0.057 \quad , \quad -0.035)} $  }& \\
			&     & Decision tree&        \multicolumn{2}{c}{${(-0.024 \quad , \quad +0.002)} $  }  &  \\
			\hline
			
			$\beta=(0.5,3.5), \lambda=5$ & & Ordinal model & \multicolumn{2}{c}{${(-0.014 \quad , \quad +0.001)} $  }  &  \\
			&     & GAMMs&       \multicolumn{2}{c}{${(-0.013\quad , \quad +0.001)} $  } &  \\
			&     & Random forest &        \multicolumn{2}{c}{${(-0.019 \quad , \quad -0.002)} $  }   & \\
			%&     & Ordinal model&        \multicolumn{2}{c}{${(-0.033 \quad , \quad -0.001)} $  } && \multicolumn{2}{c}{${(-0.025 \quad , \quad +0.007)} $  }  \\
			&     & Gradient boosting&       \multicolumn{2}{c}{${(-0.120 \quad , \quad -0.008)} $  }& \\
			&     & Decision tree&        \multicolumn{2}{c}{${(-0.031 \quad , \quad -0.014)} $  }  &  \\
			\hline
	
		\end{tabular}
	\end{small}
	\label{sim4_5}
\end{table}

\begin{table}[H]
	\caption{Summary of simulation results. $\mu_1, \cdots, \mu_6$ denote the mean error rates of HIBITS, Ordinal model (logistic regression), GAMMs, Random forest, Gradient boosting and Decision tree respectively. $1000$ simulated datasets were generated under the scenario: $\prob ({y_i}=1)=\Phi(\beta_0{x_{1i}}+\beta_1{y_{i-1}})$ (``Scenario 4''
		). We calculated the $95\%$ Bonferroni-corrected confidence intervals of the prediction error difference  from the testing dataset, $\mu_1 - \mu_i, i=2, \cdots, 6$ that the classification error rate for the proposed method is lower than that for each of competing methods.}
	\centering
	\begin{small}
		\begin{tabular}{lllccc}
			\hline
			&&& \multicolumn{2}{c}{Scenario 4} &\\
			\cline{4-6} 
			\multicolumn{2}{c}{Parameters($\beta$)}& Competing Method & \multicolumn{2}{c}{95\% confidence interval of }  & \\
			\multicolumn{2}{c}{} & &  \multicolumn{2}{c}{$\mu_1 - \mu_i, i=2, \cdots, 7$} & \\
			\hline
			
			$\beta=(0.5,3)$ & & Ordinal model & \multicolumn{2}{c}{${(-0.003 \quad , \quad +0.015)} $  }  &  \\
			&     & GAMMs&       \multicolumn{2}{c}{${(-0.006\quad , \quad +0.005)} $  } & \\
			&     & Random forest &        \multicolumn{2}{c}{${(-0.002 \quad , \quad +0.015)} $  }   & \\
			%&     & Ordinal model&        \multicolumn{2}{c}{${(-0.032 \quad , \quad -0.001)} $  } && \multicolumn{2}{c}{${(-0.006 \quad , \quad +0.003)} $  }  \\
			&     & Gradient boosting&       \multicolumn{2}{c}{${(-0.012 \quad , \quad +0.009)} $  }& \\
			&     & Decision tree&        \multicolumn{2}{c}{${(-0.016 \quad , \quad +0.008)} $  }  & \\
			\hline
			$\beta=(0.5,3.5)$ & & Ordinal model & \multicolumn{2}{c}{${(-0.003 \quad , \quad +0.007)} $  }  &  \\
			&     & GAMMs&       \multicolumn{2}{c}{${(-0.002\quad , \quad +0.008)} $  } &  \\
			&     & Random forest &        \multicolumn{2}{c}{${(-0.005 \quad , \quad +0.011)} $  }   & \\
			%&     & Ordinal model&        \multicolumn{2}{c}{${(-0.033 \quad , \quad -0.001)} $  } && \multicolumn{2}{c}{${(-0.025 \quad , \quad +0.007)} $  }  \\
			&     & Gradient boosting&       \multicolumn{2}{c}{${(-0.010 \quad , \quad +0.016)} $  }& \\
			&     & Decision tree&        \multicolumn{2}{c}{${(-0.011 \quad , \quad +0.002)} $  }  &  \\
			\hline			
			
		\end{tabular}
	\end{small}
	\label{sim4_6}
\end{table}

\begin{table}[H]
	\caption{Summary of simulation results. $1000$ simulations were generated under the scenario: $\prob ({y_i}=1)=\Phi(\beta_0{x_{1i}}+\beta_1{y_{i-1}}+\mathbf{f}(x_{2i}))$ (``Scenario 3''). We present the 95\%  confidence intervals $\beta_0$ and $\beta_1$ from the training dataset.}
	\centering
	\begin{small}
		\begin{tabular}{lllccc}
			\hline
			&&& \multicolumn{2}{c}{Scenario 3} &\\
			\cline{4-6} 
			\multicolumn{2}{c}{Parameters($\beta, \lambda$)}& Method & \multicolumn{2}{c}{95\% confidence interval of }  & \\
			\multicolumn{2}{c}{} & &  $\beta_0$ & $\beta_1$& \\
			\hline
			
			$\beta=(0.5,3), \lambda=10$ & & HIBITS method& $(0.129, 0.564)$&$(1.529, 3.574)$ & \\
			&     & Ordinal model &     $ (-0.247, 0.564)$ & $(0.756, 2.549)$&  \\
			\hline
			
			$\beta=(0.5,3), \lambda=5$ & & HIBITS method& $(0.191, 0.502)$&$(1.784, 3.956)$ & \\
			&     & Ordinal model &     $ (-0.273, 0.668)$ & $(0.966, 2.831)$&  \\
			\hline
			
			$\beta=(0.5,3.5), \lambda=10$ & & HIBITS method& $(0.129, 0.579)$&$(1.766, 3.871)$ &\\
			&     & Ordinal model &     $ (-0.406, 0.734)$ & $(0.875, 2.678)$&   \\
			\hline
			
			$\beta=(0.5,3.5), \lambda=5$ & & HIBITS method& $(0.200, 0.509)$&$(2.111, 4.310)$ & \\
			&     & Ordinal model &     $ (-0.239, 0.666)$ & $(1.156, 3.046)$&  \\
			
			\hline
		\end{tabular}
	\end{small}
	\label{sim4_7}
\end{table}

\subsection{Investigating the misspecification of the covariance function}
\label{kernel}
The objective of this section is to study the effects of misspecification on the covariance function. We will assume the true covariance function follows mixtures of different kernels and apply the proposed HIBITS method to the generated dataset. In particular, we generate binary time series $y_i$ under the following scenario:
\begin{itemize}
\item {\bf Scenario 5 (with a mixture covariance function).}\\
$\prob ({y_i}=1)=logit^{-1}(\beta_0{x_{1i}}+\beta_1{y_{i-1}}+\mathbf{f}(x_{2i})).$ 

Here, $\mathbf{f}(x_{2i}))$ follows Gaussian process with$$Cov(\mathbf{f}(x_{2i}), \mathbf{f}(x_{2j}))=\eta\big[\lambda\exp(-\rho({x_{2i}}-{x_{2j}})^2)+\sigma^2\delta_{ij}\big] + (1-\eta)\Big[\frac{1}{1 + {\tau({x_{2i}}-{x_{2j}})^2}}\Big].$$
\end{itemize}
Note that we assume the covariance function is a mixture of exponential and Cauchy kernels. This setting serves as an approach of modeling the long-term and short-term correlation on $\bm{x_2}.$ By increasing the value of trade-off parameter $\eta$, the mixture kernel will weight more on the exponential kernel, which captures the short-term dependence. Table~\ref{sim4_8} summarizes the results of mean error rates under Scenario 5. It is shown that the proposed HIBITS is able to maintain significant lower error rates compared to the other competing methods when the trade-off parameter $\eta$ is $0.2$. As we increase the value to be $0.8$, HIBITS performs almost as good as all the other methods and significantly better than decision tree. Table~\ref{sim4_9} presents the confidence intervals in Scenario 5. Similar to the previous results, HIBITS is capable of yielding narrower intervals and high capture rates even when the trade-off parameter $\eta$ is large. In summary, through this section, simulation results show that the proposed HIBITS method is robust to the misspecification of covariance function. This is partly due to the fact that we are able to dynamically ``learn" the hyperparameter through model selection. The fine-tuned covariance function could capture the long-term and short-term correlation from the generated dataset.

\begin{table}[H]
	\caption{Summary of simulation results. $\mu_1, \cdots, \mu_6$ denote the mean error rates of HIBITS, Ordinal model (logistic regression), GAMMs, Random forest, Gradient boosting and Decision tree respectively. $1000$ simulated datasets were generated under the scenario: $\prob ({y_i}=1)=logit^{-1}(\beta_0{x_{1i}}+\beta_1{y_{i-1}}+\mathbf{f}(x_{2i}))$ (``Scenario 5''). We calculated the $95\%$ Bonferroni-corrected confidence intervals of the prediction error difference  from the testing dataset, $\mu_1 - \mu_i, i=2, \cdots, 6$ that the classification error rate for the proposed method is lower than that for each of competing methods.}
	\centering
	\begin{small}
		\begin{tabular}{lllccc}
			\hline
			&&& \multicolumn{2}{c}{Scenario 5} &\\
			\cline{4-6} 
			\multicolumn{2}{c}{Parameters($\beta$, $\eta$)}& Competing Method & \multicolumn{2}{c}{95\% confidence interval of }  & \\
			\multicolumn{2}{c}{} & &  \multicolumn{2}{c}{$\mu_1 - \mu_i, i=2, \cdots, 7$} & \\
			\hline
			
			$\beta=(0.5,3), \eta=0.2$ & & Ordinal model & \multicolumn{2}{c}{${(-0.024 \quad , \quad -0.011)} $  }  &  \\
			&     & GAMMs&       \multicolumn{2}{c}{${(-0.023\quad , \quad -0.010)} $  } & \\
			&     & Random forest &        \multicolumn{2}{c}{${(-0.022 \quad , \quad -0.002)} $  }   & \\
			%&     & Ordinal model&        \multicolumn{2}{c}{${(-0.032 \quad , \quad -0.001)} $  } && \multicolumn{2}{c}{${(-0.006 \quad , \quad +0.003)} $  }  \\
			&     & Gradient boosting&       \multicolumn{2}{c}{${(-0.038 \quad , \quad -0.023)} $  }& \\
			&     & Decision tree&        \multicolumn{2}{c}{${(-0.061 \quad , \quad -0.037)} $  }  & \\
			\hline
			$\beta=(0.5,3), \eta=0.8$ & & Ordinal model & \multicolumn{2}{c}{${(-0.008 \quad , \quad +0.005)} $  }  &  \\
			&     & GAMMs&       \multicolumn{2}{c}{${(-0.008\quad , \quad +0.007)} $  } &  \\
			&     & Random forest &        \multicolumn{2}{c}{${(-0.005 \quad , \quad +0.001)} $  }   & \\
			%&     & Ordinal model&        \multicolumn{2}{c}{${(-0.033 \quad , \quad -0.001)} $  } && \multicolumn{2}{c}{${(-0.025 \quad , \quad +0.007)} $  }  \\
			&     & Gradient boosting&       \multicolumn{2}{c}{${(-0.004 \quad , \quad +0.001)} $  }& \\
			&     & Decision tree&        \multicolumn{2}{c}{${(-0.016 \quad , \quad -0.002)} $  }  &  \\
			\hline			
			
		\end{tabular}
	\end{small}
	\label{sim4_8}
\end{table}

\begin{table}[H]
	\caption{Summary of simulation results. $1000$ simulations were generated under the scenario: $\prob ({y_i}=1)=logit^{-1}(\beta_0{x_{1i}}+\beta_1{y_{i-1}}+\mathbf{f}(x_{2i}))$ (``Scenario 5'').We present the 95\%  confidence intervals $\beta_0$ and $\beta_1$ from the training dataset.}
	\centering
	\begin{small}
		\begin{tabular}{lllccc}
			\hline
			&&& \multicolumn{2}{c}{Scenario 5} &\\
			\cline{4-6} 
			\multicolumn{2}{c}{Parameters($\beta, \eta$)}& Method & \multicolumn{2}{c}{95\% confidence interval of }  & \\
			\multicolumn{2}{c}{} & &  $\beta_0$ & $\beta_1$& \\
			\hline
			
			$\beta=(0.5,3), \eta=0.2$ & & HIBITS method& $(0.056, 0.505)$&$(1.796, 3.306)$ & \\
			&     & Ordinal model &     $ (-0.232, 0.671)$ & $(0.829, 2.769)$&  \\
			\hline
			
			$\beta=(0.5,3), \eta=0.8$ & & HIBITS method& $(0.160, 0.702)$&$(2.803, 3.309)$ & \\
			&     & Ordinal model &     $ (-0.333, 1.142)$ & $(1.186, 6.428)$&  \\
			\hline

		\end{tabular}
	\end{small}
	\label{sim4_9}
\end{table}

\section{Analysis of the sleep state data}
\label{realdata}
In this section, we apply our method to sleep state data. People spend one third of their lifetime on sleep. Studying and predicting sleep patterns is significant because our body requires sleep in much the same way as the need of eating and breathing. Moreover, disruptions in sleep are known 
to be associated with both psychiatric and chronic diseases. 
%During our sleep, the energy is reserved and the basal metabolism is decreased by 5-25\% \citep{Nevsimalova:1997}. NREM sleep tends to be involved in disturbance of emotional behavior like concentration and interest for distinct goal \citep{Cai:1989}. During REM sleep, the maturation of brain and myelinization of nerve fibers proceed \citep{Nevsimalova:1997}. Moreover,  \citet{Stickgold:1998} found that both of REM and NREM sleeps are significant in memory consolidation and learning. Particularly in REM sleep, the insignificant bindings are abolished and the memory traces are fixed \citep{Susmakova:2005}. 
In what follows, we will analyze the sleep data obtained from an observational study with the 
goal predicting sleep states and identifying associations between sleep states and potential 
regulators such as temperature and heart rate. 

\subsection{Exploratory analysis}
The data were recorded from a four month old infant who was placed to bed at night. Heart rate ($H_i$, beats per minute at time $i$), temperature ($T_i$, in Celsius, at time $i$) and sleep stage ($S_i$ at time $i$) of length ($N = 1024$) were sampled every 30 seconds. Heart rate was recorded automatically using a standard ECG (electrocardiogram) monitor. The infant's EEG (electroencephalogram) and EOG (electrooculogram) were also measured with a period of 30 seconds. The EEG captured brain waves including alpha (8 -- 15 Hz), beta (16 -- 31 Hz) and mu (8 -- 12 Hz) rhythms; EOG recorded the eye movement. Sleep stage for each time point $i$ was determined by the sleep lab expert visually interpreting the EEG and EOG record \citep{Nevsimalova:1997}. It was classified as 4 categories: (1) quiet sleep, (2) indeterminate sleep, (3) active sleep and (4) awake \citep{benbadis2006introduction}. The sleep stage $S_i$ was measured as integers ranging from 1 to 4. In this section, following the work of \citet{fokianos2002regression}, sleep state is defined 
as a binary time series $Y_i$: 
\begin{displaymath}
   Y_i = \left\{
     \begin{array}{ll}
       1 & : \text{awake at time $i$},\\
       0 & : \text{not awake at time $i$}.
     \end{array}
   \right.
\end{displaymath}
where ``not awake'' stands for quiet sleep, indeterminate sleep or active sleep.

%Table \ref{ds2} presents the summary of descriptive statistics of the covariates. It can be found that the heart rate is 134.13 beats per minute on average with standard deviation 15.14. Temperature has an average of 37.13 Celsius and standard deviation of 0.15. 

\begin{figure}[th!] \centering
                            \begin{tabular}{cc}
                            \includegraphics[width=.5\textwidth]{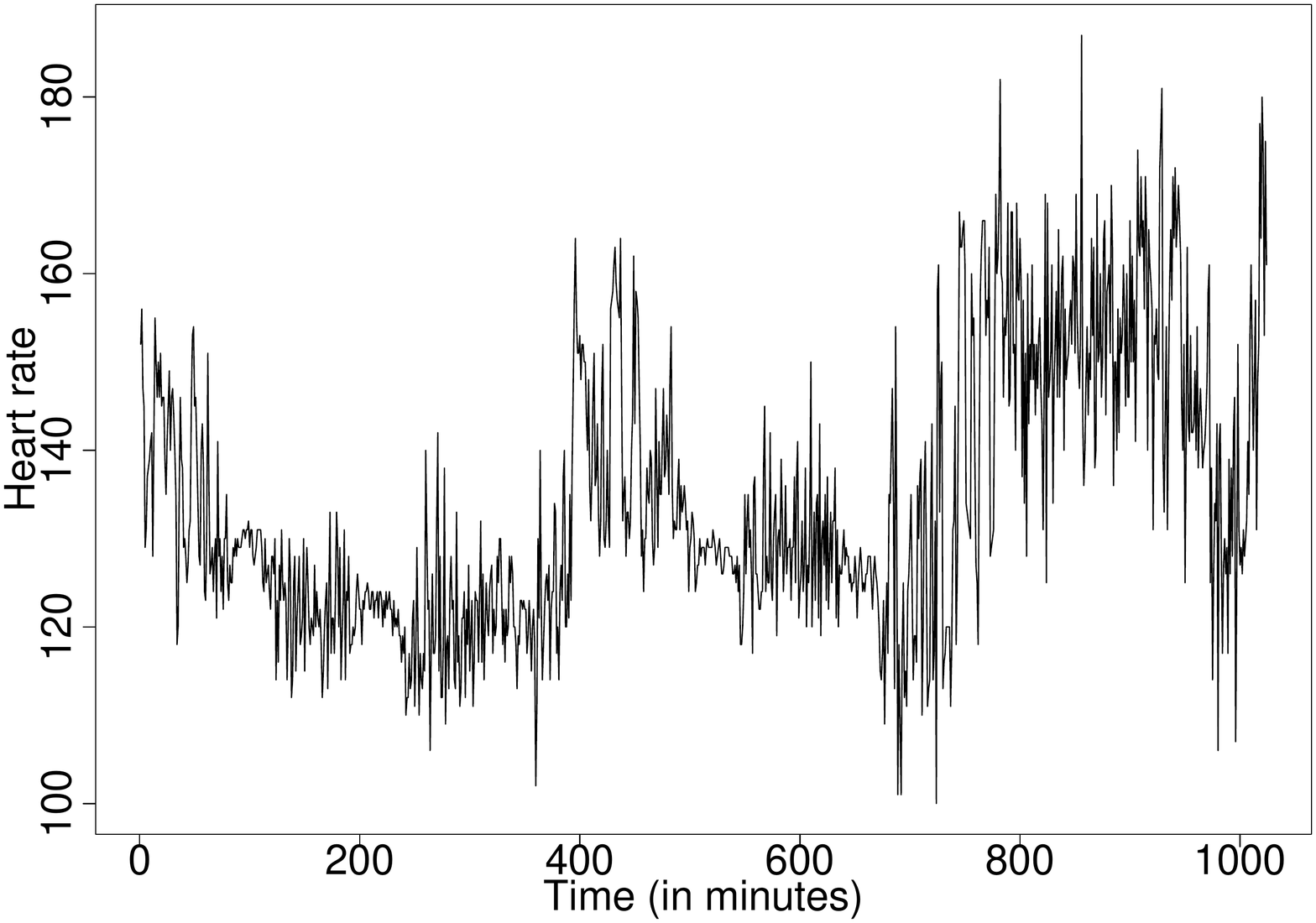} &
                            \includegraphics[width=.5\textwidth]{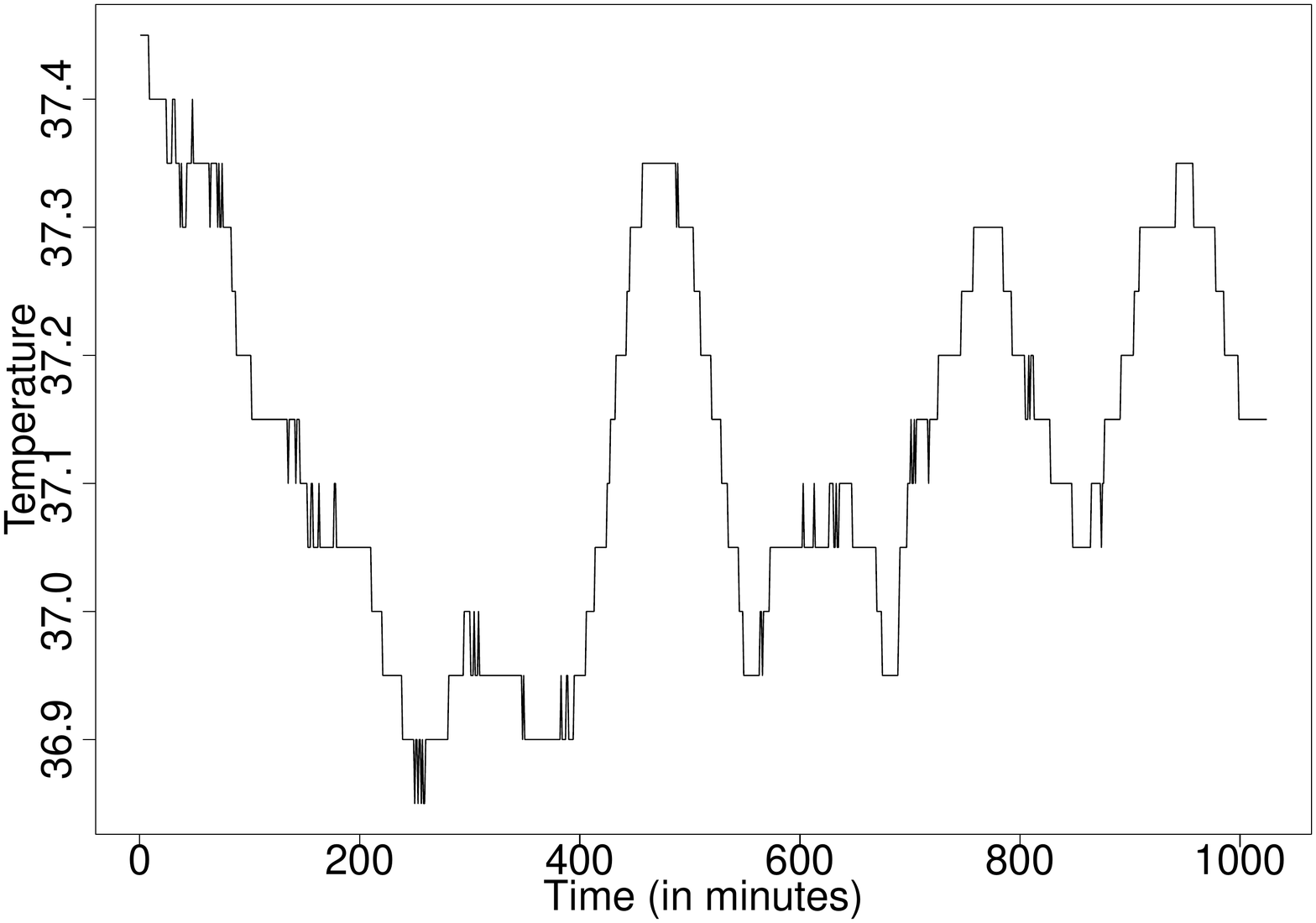} \\                
                             Heart rate time series& Temperature time series
                            \end{tabular}
                            \caption{Left: heart rate (in beats per minute). Right: 
                            temperature (in Celsius).}
                            \label{expl1}
\end{figure}

Time series plots of heart rate, temperature and sleep state are shown in Figure \ref{expl2} and Figure \ref{expl1}. By comparing the heart rate, temperature with sleep state, we note that 
higher heart rate are likely to correspond to sleep state 1 (awake).  While this pattern is clear 
for heart rate,  no such pattern between temperature and sleep state can be detected by 
visual inspection. In addition, it can be seen that the current sleep state is highly related to previous states. 

\begin{figure}[th!] \centering
	\begin{tabular}{cc}
		\includegraphics[width=.5\textwidth]{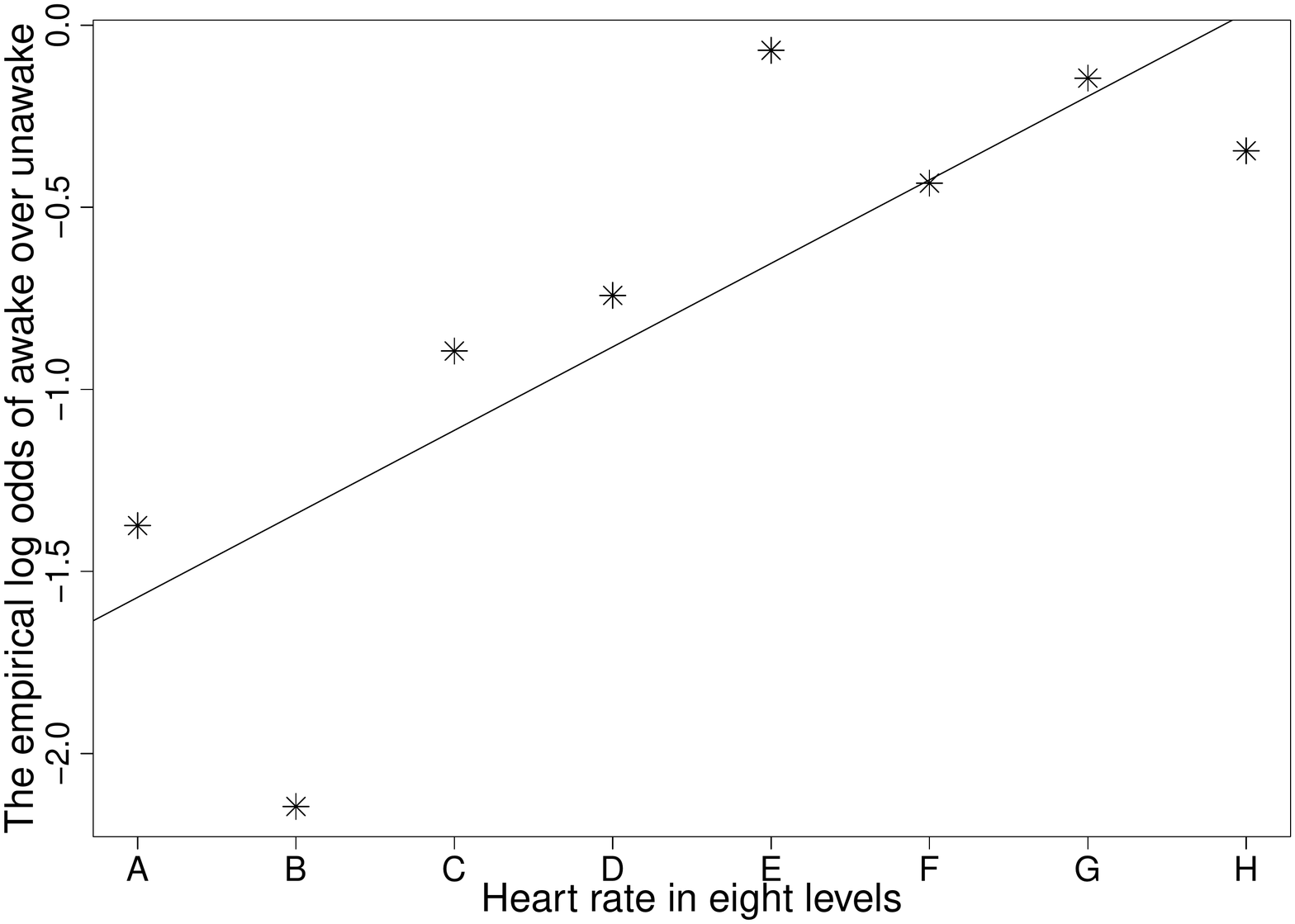} &
		\includegraphics[width=.5\textwidth]{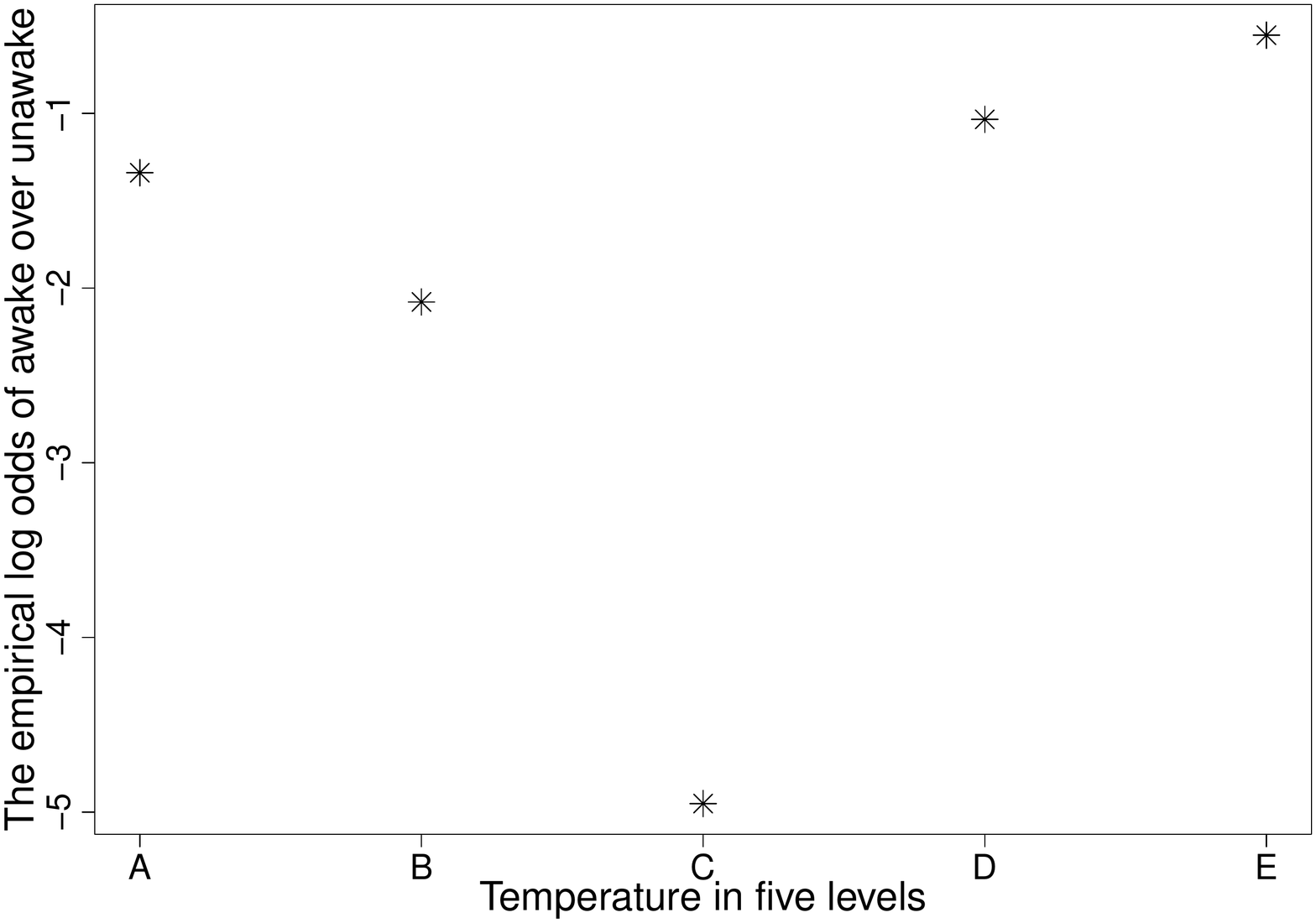} \\
		Empirical log odds versus heart rate &  Empirical log odds versus temperature
	\end{tabular}
	\caption{Scatterplots of empirical log odds versus heart rate and temperature. The left panel shows the empirical log odds over eight levels of heart rate. The right panel displays the same value versus temperature.}
	\label{heart_temp}
\end{figure}

To further study the dependence of sleep state on the covariates temperature and heart rate, we conducted additional preliminary analysis. Particularly, we categorize heart rate and temperature (after taking the logarithm) into several levels and calculate the empirical log odds of awake over not-awake for each level. Figure \ref{heart_temp} show the relationship between the empirical log odds and different levels of the underlined heart rate and temperature.  We are able to identify a positive association between heart rate with current sleep states. The effect of the lower heart rates are associated with higher probability of being asleep. Regarding temperature, one 
can hardly identify any definitive relationship using the log odds. Moreover, in Table \ref{transit}, we report 
the empirical transition probability of sleep state. It shows that the current sleep state is highly dependent on the previous state.  More specifically, 
there is a strong tendency for sleep to remain in its current state.

%comments start here
\iffalse
\begin{table}[h]
\centering
\caption{Descriptive statistics of variables}
\label{ds2}
\begin{tabular}{lllll}
\hline
Variables & Mean & Standard deviation & Range & Skewness \\ \hline
      Heart rate (beats per minute)   &  134.13    &    15.14               &   87.00   &  0.72\\
       Temperature (Celsius)   &  37.13   &    0.15              &  0.60  & 0.13 \\
 \hline
\end{tabular}
\end{table}
\fi

\begin{table}[h]
\centering
\caption{Empirical transition table of sleep state: when the current state is not awake, the
sample probability of staying not wake in the next time point is $729/735$ while the sample
probability of being in the awake state at the next time point is $6/735$. When the current 
state is awake, the sample probability of staying awake at the next time point is $282/288$ 
while the sample probability of changing to a non-awake state at the next time point is 
$6/288$. }
\label{transit}
\begin{tabular}{ccc}
\hline
 & $Y_ {i-1}=0$& $Y_ {i-1}=1$ \\ \hline
$Y_ {i}=0$ &     729/1023      & 6/1023  \\ \hline
$Y_ {i}=1$ &    6/1023       & 282/1023  \\ \hline
\end{tabular}
\end{table}

\subsection{Modeling and results}
Following the exploratory analysis, $\log H_i$, $Y_{i-1}$ and time (in minutes) are suggested in the proposed model. Since there is strong effect of $\log H_i$, $Y_{i-1}$ on current sleep state, we include those two covariates as fixed effect components. Gaussian process on time domain is introduced to capture the nonlinear term. 

\begin{figure}[th!] \centering
                           % \begin{tabular}{cc}
                            \includegraphics[width=.6\textwidth]{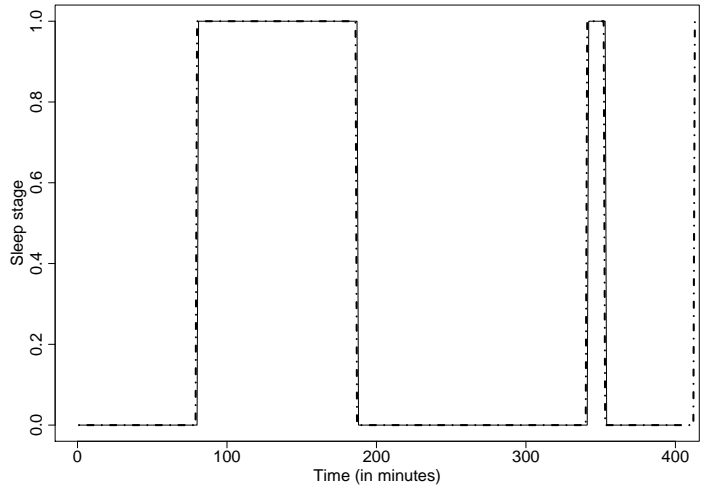} %&
                                                     \caption{Predicted sleep state (solid line) overlaid with real data (dotted line) (training/testing data size 600/400).}
                            \label{real1}
\end{figure}

We also applied our proposed binary hybrid approach to make the inference and prediction. Summaries of point and interval estimates are shown in Table \ref{table_real}. It is seen that compared with the ordinal model (logistic regression), the point estimates are similar.  However, there is 
significantly large difference among the interval estimates. Using the proposed HIBITS method, we gain substantially narrower confidence intervals than ordinal model. The benefits are up to almost $90\%$ shorter in length. From the proposed results, we find that one unit increment in heart rate at current time point will lead to $211.4\%$ accretion of odds. Current odd of sleep state when previous sleep state is awake is estimated to be dramatically higher than that when previous state is not awake. To test the prediction power of this model, the proposed method was implemented with various training and testing data size. Numerical results are summarized in Table \ref{pt1}. It can be seen that the model produces around 99\% prediction accuracy while ordinal model yields about 96\%. As we decrease the ratio of training over testing data size, the prediction accuracy remains stable. Time series plots of the real and predicted sleep state are presented in Figure \ref{real1}. It can be shown that the proposed method produces high prediction accuracy and recover the same sleep state pattern as the real dataset. To check for the sensitivity of the proposed method to the estimated value of parameter $\lambda$, we compared the results from the data-adaptive estimate ($0.730$) against the following values ($1.730, 2.730$). The data-adaptive estimate gave roughly the same prediction error but the confidence intervals were narrower.

\begin{table}[h]
	\caption{Summary of the sleep state analysis. The point and interval estimates from HIBITS method are obtained by Section~\ref{infer}. It can be seen that the widths of the confidence intervals from the HIBITS method are narrower than those of the classical ordinal model.}
	\begin{small}
		\hskip1.3cm
		\begin{tabular}{lllccc}
			\hline
			\multicolumn{2}{c}{Parameters($\beta_0, \beta_1$)}& Method & Point estimate &{95\% confidence intervals}  & \\
			\hline
			
			$\beta_0$ & & HIBITS method& $1.136$&$(1.000, 1.271)$ &\\
			&     & Ordinal model &     $ 1.105$ & $(0.101, 2.109)$&  \\
			\hline
				$\beta_1$ & & HIBITS method& $8.275$&$(8.124, 8.427)$ &\\
				&     & Ordinal model &     $ 8.241$ & $(6.669, 9.813)$&  \\

			\hline
		\end{tabular}
	\end{small}
	\label{table_real}
\end{table}

\begin{table}[h]
\centering
\caption{Prediction accuracy with different training and testing data size.}
\label{pt1}
\begin{tabular}{llc}
\hline
Training/Testing data size& Method &Prediction Accuracy \\ \hline
 600/400 &HIBITS method& $99.0\%$ \\ 
  &Ordinal model& $96.0\%$ \\ 
 \hline
500/500 &HIBITS method& $99.2\%$ \\
&Ordinal model& $96.1\%$ \\ 
\hline
400/600 & HIBITS method&$99.1\%$ \\ 
 &Ordinal model& $96.4\%$ \\ 
\hline
\end{tabular}
\end{table}

\subsection{Discussion on missing data}
One advantage of the proposed prediction model is that it captures the information from its own past. Derived from the results, the odds when previous sleep state is awake is 4000-fold higher than that when the preceding state is not awake. However, if there are missing data or the observations are not collected successively, such information will be lost. This motivates us to adjust the model to fit such cases. In the adjusted model, we choose $\log H_i$ as fixed effects and still use Gaussian process on time domain. To test the prediction power, instead of fixing the training and testing dataset, we randomly pick those two pieces of data with fixed size. The proposed HIBITS method was implemented. Summaries of the test results can be found in Table \ref{pt}. The tests were conducted 10 times with training and testing data of different sizes. From the results, it is clear that as the training data size becomes larger, the prediction accuracy increases at a reasonable rate. As the training data size reaches 600, the accuracy is promising.
\begin{table}[h]
\caption{Prediction accuracy with different training and testing data size, *stands for the test number.}
\label{pt}
\centering
\begin{tabular}{llllllllllll}
\hline
Training/Testing & 1* &  2 &  3&  4 & 5&  6&  7&  8&  9 & 10 & Average\\
data size  &&&&&&&&&&& \\ \hline
 400/100 & 0.76 & 0.92 & 0.95 & 0.93 & 0.80 & 0.64 & 0.84 & 0.89 & 0.90 &  0.93 & 0.86 \\
 \hline
 500/100 & 0.99 & 0.89 & 0.90 & 0.98 & 0.91 & 0.88 & 0.89 & 0.93 & 0.95 &  0.91 & 0.92 \\ 
 \hline
600/100 & 0.95 & 0.97 & 0.99 & 0.91 & 0.93 & 0.96 & 0.97 & 0.90 & 0.92 & 0.93 & 0.94 \\ \hline
\end{tabular}
\end{table}

To further study the performance of the proposed HIBITS method, we make the ratio of training over testing data size smaller. Particularly, we change the training and testing data size to be 700 and 300 respectively. The prediction accuracy is around 0.873. If we move further to change the training data size to be 800 and testing data size to be 200, the prediction accuracy is about $90\%$. All the results demonstrate that the proposed binary hybrid method produces promising prediction power when the dataset are not collected successively or partly missing. Moreover, it should be pointed out that the computation is not very demanding. The tests are conducted in R programming and the operation time is approximately 90 seconds for each test. 

\section{Concluding remarks}
The proposed hybrid inference method for binary time series (HIBITS) produces efficient inference and promising predictions with a relatively low computational cost. Compared to existing methods, our proposed approach has the following advantages: on one hand, by involving known covariates as fixed effect components, we make use of the information indicating the association between the response and covariates. On the second stage, a Gaussian process will capture the information beyond what provided by those covariates of both endogenous and exogenous time series. On the other hand, as indicated in the simulation, the proposed method is robust compared to existing methods. The proposed model selection strategy allows the model to fit the data even though not enough information is captured by the fixed effect components. The strategies in providing point and interval estimates, in addition, allows researchers to gain more informative conclusions in the association between response and covariates. These advantages make our model easy to interpret. In summary, the proposed HIBITS method, serving as an approach with high prediction power, efficient inference capability and direct interpretability, provides a  promising methodology in modeling and predicting sleep states and other binary time series.  

Although the HIBITS method produced outstanding results, it is limited to binary outcomes. As a future direction, we could extend the proposed HIBITS to model general categorical responses (e.g. the 4 sleep stages). In the case of nominal categorical outcomes, we could follow the similar framework of multinomial logit model discussed by \cite{fokianos2002regression}. Specifically, the link function could be extended to softmax function where fixed and random effects can be imposed on the systematic component, which is a natural generalization of our proposed HIBITS. On the other hand, if the outcomes are ordinal categorical time series, one can impose a threshold mechanism on the systematic component of the model. Following the scenario of proportional odds models in \cite{fokianos2003regression}, the HIBITS method
can be extended by incorporating fixed and random effects. 

\section*{Acknowledgement}
The authors thank the anonymous reviewers for providing insightful comments and suggestions. This work was supported in part by grants awards to H. Ombao
(NSF DMS 1509023 and NSF MMS 1461543) and B. Shahbaba 
(NIH R01-AI107034 and NSF DMS-1622490). 

\bibliographystyle{chicago}
\bibliography{xg}
\newpage
\appendix
\section{Manual of R package {\tt HIBITS}}
\subsection{Description}
\textbf{Package}: HIBITS\\
\textbf{Type}: Package\\
\textbf{Title}: The hybrid inference method of binary time series\\
\textbf{Version}: 0.1\\
\textbf{Date}: 2016-03-23\\
%\textbf{Authors@R}: person(``Xu", ``Gao", email =``xgao2@uci.edu", role = c(``aut", ``cre"))\\
\textbf{Description}: Generate the simulated binary time series, compute the point and
confidence intervals of the parameters in the model and make predictions
of the future observations\\
\textbf{License}: GPL-2\\
\textbf{LazyData}: TRUE\\
\textbf{RoxygenNote}: 5.0.1\\
\textbf{Imports}: stats, base\\
\textbf{NeedsCompilation}: no\\
%\textbf{Packaged}: 2016-03-24 05:43:58 UTC; xugao\\
%\textbf{Author}: Xu Gao [aut, cre]\\
%\textbf{Maintainer}: Xu Gao $<$xgao2@uci.edu$>$\\
\textbf{Built}: R 3.2.3; ; 2016-03-24 05:54:32 UTC; unix
\subsection{Functions}
\subsubsection{BHM}
\textbf{Description} The implement of the hybrid method\\
\textbf{Usage} BHM(xtrain, ytrain, xtest, rho, sigma, n, train\_index, test\_index, optim = F, specify, ini\_interval)\\
\textbf{Argument}\\
\begin{tabular}{ll}
xtrain&The training datasets of features\\
ytrain& The training datasets of response\\
xtest & The testing datasets of features\\
rho & The hyperparameter of the kernel function\\
sigma & The hyperparameter of the kernel function\\
n& The size of the whole dataset\\
train\_index& The index of the training dataset\\
test\_index&	The index of the testing dataset\\
optim& An option of using the model selection strategy\\
specify & Pre-specified value of parameters if optim is False\\
ini\_interval & The range of lambda
\end{tabular}
\newpage
\noindent \textbf{Value}\\
BHM \quad A list of ``predictions", ``the estimated parameters", ``the estimated intercept from the maximum likelihood " and ``fhat".\\
\textbf{Examples}\\
x1 = rnorm(150);\\
time\_series = gen\_sim(rho = 1, sigma = 0.1, lambda = 5, beta\_true = c(0.5, 2), x1, length(x1), ini = 1, F)\$realization\\
split\_data\_temp = Split\_data(cbind(x1, time\_series), 100);\\
temp = BHM(split\_data\_temp\$xtrain, split\_data\_temp\$ytrain, split\_data\_temp\$xtest, rho = 1, sigma = 0.1, length(x1), split\_data\_temp\$train\_index,
split\_data\_temp\$test\_index, T, 5, c(0,10))

\subsubsection{gen\_sim}
\textbf{Description} Generate binary time series for simulation purpose\\
\textbf{Usage} gen\_sim(rho, sigma, lambda, beta\_true, x, n, ini, GPs = T)\\
\textbf{Argument}\\
\begin{tabular}{ll}
	rho	&
	The hyerparameter of the kernel function\\
	sigma	&
	The hyerparameter of the kernel function\\
	lambda	&
	The parameter of the kernel function\\
	beta\_true &	
	The parameters of the systematic components\\
	x	&
	The covariates to generate time series\\
	n	&
	The size of the whole dataset\\
	ini	&
	The initial value of the time series\\
	GPs	&
	An option of involving a Gaussian process in generating the time series\\
	&
\end{tabular}\\
\textbf{Value}\\
gen\_sim \quad  A list of ``realizations" of time series and the ``Kernel" of the covariance matrix.
\textbf{Examples}\\
x1 = rnorm(150);\\
time\_series = gen\_sim(rho = 1, sigma = 0.1, lambda = 5, beta\_true = c(0.5, 2), x1, length(x1), ini = 1, F)\$realization
\subsubsection{interval\_est }
\textbf{Description} Generate the interval estimates derived from the hybrid method\\
\textbf{Usage} interval\_est(xtrain, ytrain, rho, sigma, lambda, beta, intercept, fhat)\\
\newpage
\noindent \textbf{Argument}\\
\begin{tabular}{ll}
	xtrain	&
	The training dataset of features\\
	ytrain	&
	The training dataset of response\\
	rho	&
	The hyerparameter of the kernel function\\
	sigma	&
	The hyerparameter of the kernel function\\
	lambda	&
	The estimated parameter of the kernel function\\
	beta	&
	The estimated parameters of the systematic component\\
	intercept	&
	The estimated intercept derived from the maximum likelihood\\
	fhat & Value obtained from the estimated Gaussian process.\\
	&\\
\end{tabular}\\
\textbf{Value}\\
interval\_est  \quad A list of samples of paramters from which we can get the interval estimates.
\textbf{Examples}\\
x1 = rnorm(150);\\
time\_series = gen\_sim(rho = 1, sigma = 0.1, lambda = 5, beta\_true = c(0.5, 2), x1, length(x1), ini = 1, F)\$realization;\\
split\_data\_temp = Split\_data(cbind(x1, time\_series), 100);\\
temp = BHM(split\_data\_temp\$xtrain, split\_data\_temp\$ytrain, split\_data\_temp\$xtest, rho = 1, sigma = 0.1, length(x1), split\_data\_temp\$train\_index,
split\_data\_temp\$test\_index, T, 5, c(0, 10));\\
para\_sample = interval\_est(split\_data\_temp\$xtrain, split\_data\_temp\$ytrain, rho = 1, sigma = 0.1, temp\$parameter[3], temp\$parameter[-3], temp\$intercept, temp\$fhat)\$par
\subsubsection{interval\_glm}
\textbf{Description} Generate the interval estimates derived from logistic regression (ordinal model)\\
\textbf{Usage} interval\_glm(xtrain, ytrain)\\
\textbf{Argument}\\
\begin{tabular}{ll}
xtrain	&
The training dataset of features\\
ytrain	&
The training dataset of response\\
&\\
\end{tabular}\\
\textbf{Value}\\
interval\_glm \quad  A list of confidence intervals of parameters from logistic regression (ordinal model).\\
\textbf{Examples}\\
x1 = rnorm(150);\\
time\_series = gen\_sim(rho = 1, sigma = 0.1, lambda = 5, beta\_true = c(0.5, 2), x1, length(x1), ini = 1, F)\$realization\\
split\_data\_temp = Split\_data(cbind(x1, time\_series), 100);\\
para\_sample = interval\_glm(split\_data\_temp\$xtrain, split\_data\_temp\$ytrain)
\subsubsection{Split\_data}
\textbf{Description} Split the dataset into training and testing dataset\\
\textbf{Usage} Split\_data(dataset, ntrain)\\
\textbf{Argument}\\
\begin{tabular}{ll}
	dataset	&
	The whole dataset.\\
	ntrain	&
	The size of training dataset.\\
	& \\
\end{tabular}\\
\textbf{Value}\\
Split\_data \quad  A list of ``feature training data", ``response training data", ``feature testing data", ``response testing data", ``training data index", ``testing data index".\\
\textbf{Examples}\\
demo = replicate(5, rnorm(200));\\
Split\_data(demo, 150);\\
\subsubsection{like}
\textbf{Description} Calculate the value of logistic function\\
\textbf{Usage} like(f)\\
\textbf{Argument}\\
\begin{tabular}{ll}
f	& 
A number or a vector.\\
\end{tabular}\\
\textbf{Value}\\
The value of logistic function of f\\
\textbf{Examples}\\
like(0.5)\\
like(c(1, 3))
\end{document}